\documentclass[%
 reprint,
 superscriptaddress,
 amsmath,amssymb,
 prl,
]{revtex4-2}

\usepackage{graphicx}
\usepackage{dcolumn}
\usepackage{bm}

\usepackage{newtxtext,newtxmath}

\begin{document}

\title{Strongly Coupled Spin Waves and Surface Acoustic Waves at Room Temperature}

\author{Yunyoung Hwang}
\affiliation{%
 Institute for Solid State Physics, University of Tokyo, Kashiwa 277-8581, Japan
}%
\affiliation{%
 CEMS, RIKEN, 2-1, Hirosawa, Wako 351-0198, Japan
}%

\author{Jorge Puebla}%
 \email{jorgeluis.pueblanunez@riken.jp}
\affiliation{%
 CEMS, RIKEN, 2-1, Hirosawa, Wako 351-0198, Japan
}%

\author{Kouta Kondou}
\affiliation{%
 CEMS, RIKEN, 2-1, Hirosawa, Wako 351-0198, Japan
}%

\author{Carlos Gonzalez-Ballestero}
\affiliation{%
 Institute for Quantum Optics and Quantum Information of the Austrian Academy of Sciences, A-6020 Innsbruck, Austria
}%
\affiliation{%
 Institute for Theoretical Physics, University of Innsbruck, A-6020 Innsbruck, Austria
}%

\author{Hironari Isshiki}
\affiliation{%
 Institute for Solid State Physics, University of Tokyo, Kashiwa 277-8581, Japan
}%

\author{Carlos S\'{a}nchez Mu\~{n}oz}
\affiliation{%
 Departamento de F\'{i}sica Te\'{o}rica de la Materia Condensada and Condensed Matter Physics Center (IFIMAC), Universidad Aut\'{o}noma de Madrid, 28049 Madrid, Spain
}%

\author{Liyang Liao}
\affiliation{%
 Institute for Solid State Physics, University of Tokyo, Kashiwa 277-8581, Japan
}%

\author{Fa Chen}
\affiliation{%
 School of Integrated Circuits, Wuhan National Laboratory for
Optoelectronics, Huazhong University of Science and Technology, Wuhan 430074,
People’s Republic of China.
}%

\author{Wei Luo}
\affiliation{%
 School of Integrated Circuits, Wuhan National Laboratory for
Optoelectronics, Huazhong University of Science and Technology, Wuhan 430074,
People’s Republic of China.
}%

\author{Sadamichi Maekawa}
\affiliation{%
 CEMS, RIKEN, 2-1, Hirosawa, Wako 351-0198, Japan
}%
\affiliation{%
 Advanced Science Research Center, Japan Atomic Energy Agency, Tokai 319-1195, Japan
}%
\affiliation{%
 Kavli Institute for Theoretical Sciences, University of Chinese Academy of Sciences, Beijing 100049, People’s Republic of China
}%

\author{Yoshichika Otani}
\email{yotani@issp.u-tokyo.ac.jp}
\affiliation{%
 Institute for Solid State Physics, University of Tokyo, Kashiwa 277-8581, Japan
}%
\affiliation{%
 CEMS, RIKEN, 2-1, Hirosawa, Wako 351-0198, Japan
}%

\date{\today}

\begin{abstract}
Here, we report the observation of strong coupling between magnons and surface acoustic wave (SAW) phonons in a thin CoFeB film constructed in an on-chip SAW resonator by analyzing SAW phonon dispersion anticrossings.
Our device design provides the tunability of the film thickness with a fixed phonon wavelength, which is a departure from the conventional approach in strong magnon--phonon coupling research.
We detect a monotonic increase in the coupling strength by expanding the film thickness, which agrees with our theoretical model.
Our work offers a significant way to advance fundamental research and the development of devices based on magnon--phonon hybrid quasiparticles.
\end{abstract}

\maketitle
Hybridization between two systems can be characterized by a comparison of the coupling strength $g$ and the relaxation rates of each system $\kappa_1$ and $\kappa_2$. When $g / {\mathrm{max}}(\kappa_1, \kappa_2) > 1$, the hybridized state is in the strong coupling regime~\cite{Kockum2019}.
In the case of magnon–phonon coupling, reducing the magnon relaxation rate is experimentally challenging, thus increasing $g$ is the most efficient route to realize strong coupling. The most straightforward approach to achieve higher coupling strength is increasing the number of spins coupled in phase to the desired mode~\cite{Kockum2019}.
This approach is typically done for magnons coupling to photons in cavity magnonics experiments by just increasing the volume of the magnet~\cite{Tabuchi2014}.
The rationale behind this approach lies in the relatively spatially homogeneous microwave cavity modes throughout the magnet, especially for magnets significantly smaller than the microwave wavelength ($\sim$ mm).

However, in the context of magnon–phonon coupling research, this approach is not always straightforward as the magnon and phonon are not necessarily in phase across the sample, and thus the magnon–phonon coupling depends non-trivially on the sample geometry. For instance, in a spherical magnet, the magnon--phonon coupling decreases with the volume~\cite{Zhang2016,CGBPRL,CGBPRB}.
Nonetheless, one can circumvent this apparent limitation by choosing an appropriate geometry for the magnetic medium: a thin film with a much smaller thickness than the involved acoustic wavelength. In this system and regime, the coupling strength recovers its characteristic monotonic increase with the expansion of the film thickness, and hence, the thin film limit allows us to explore and systematically characterize the dependence of magnon–phonon coupling on the number of spins. In order to make it possible, it is crucial to attain independent control over the phonon wavelengths and the geometry of the magnetic material, thereby enabling the realization of the thin film limit and promoting high magnon–phonon coupling. One viable approach to accomplish this limit is injecting surface acoustic waves (SAWs) with variable wavelengths into the magnetic material. However, while there have been significant research on the coupling between magnons and SAW phonons over the past two decades~\cite{Weiler2011,Dreher2012,Weiler2012,Labanowski2016,Li2017,Xu2018,Hwang2020,Xu2020,Babu2021,Hatanaka2022,Gao2022,Hwang2022}, no study has reported the observation of strong coupling between magnons and SAW phonons.

\begin{figure}[]
\centering
\includegraphics{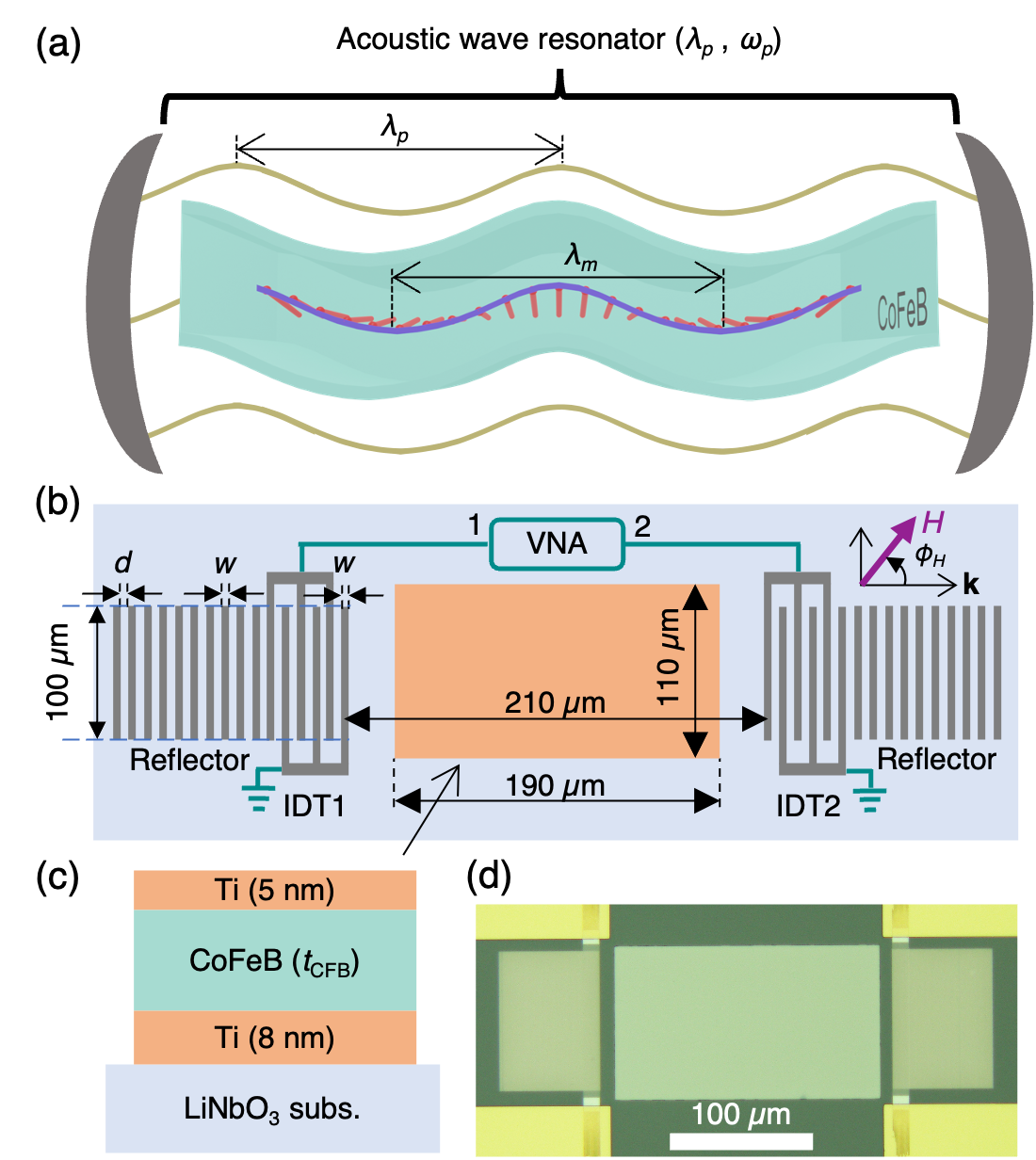}
\caption{\label{fig1} (Color online)
(a) Schematic illustration of strong magnon–phonon coupling
in an acoustic cavity that confines phonons with a wavelength $\lambda_p$ and a frequency $\omega_p$.
The yellow curves denote acoustic waves (phonons) created within the acoustic cavity and the red arrows represent magnetization dynamics.
The phonons propagating through a CFB thin film excite a spin wave (magnon),
which is represented by the purple curve, with a matched wavelength $\lambda_m = \lambda_p$.
(b) Schematic top view of the device structure used in this research.
IDT1 and IDT2 include 20 pairs of Al stripes and each set of reflectors includes 200 Al stripes.
The gap ($d$) and the width ($w$) of one Al stripe of IDTs and reflector gratings are both 150 nm.
IDT1 and IDT2 are respectively connected to port1 and port2 of a vector network analyzer (VNA).
(c) Schematic side view of the Ti/CFB/Ti stack.
(d) Optical microscope image of one of the devices used in this study.}
\end{figure}

In this Letter, we demonstrate for the first time the coupling between magnons and SAW phonons caused by SAW-driven spin wave resonance (SWR) in a Co\textsubscript{20}Fe\textsubscript{60}B\textsubscript{20} (CFB) thin film on a LiNbO\textsubscript{3} substrate at room temperature by measuring anticrossing of SAW phonon dispersion, an indication of strong interaction~\cite{Huebl2013}.
We generate SAWs from outside the CFB film, which possesses a high magnetoelastic coupling coefficient, using a high-frequency two-port SAW resonator that consists of SAW generation and detection devices enclosed by distributed Bragg reflector-like stripes, forming an acoustic cavity~\cite{Bell1976,Hwang2020,Hatanaka2022,Hwang2022,Puebla2022AnnPhys,Hwang2023},
similar to cavity quantum electrodynamics experiments. This device design allows driving acoustic modes with any wavelength $\lambda_p$ (frequency $\omega_p$) enabling coupling to magnons at the wavelength $\lambda_m = \lambda_p$ by SWR, as depicted in Fig.~\ref{fig1}(a), circumventing the reliance of $\lambda_p$ on the material geometry.
By harnessing this breakthrough, we successfully estimated the magnon--phonon coupling strength ($g$ values) of samples with the same device structure but varying CFB thicknesses ($t_{\mathrm{CFB}}$) by fitting the observed anticrossing with our theoretical model.
Our systematic study reveals a monotonic increase in $g$ with increasing $t_{\mathrm{CFB}}$, which agrees with our expectation and our magnon–phonon coupling model. This increase of $g$ allowed us to achieve strong coupling; $g / {\mathrm{max}}(\kappa_m, \kappa_p) > 1$, where $\kappa_{m(p)}$ is the magnon (phonon) relaxation rate, for devices with $t_{\mathrm{CFB}} \geq$ 20 nm.

To generate SAWs, we utilized interdigital transducers (IDT)
that can generate and detect SAWs on a piezoelectric substrate~\cite{Bell1976}.
We fabricated acoustic cavity devices including Ti (8 nm) / CFB ($t_{\mathrm{CFB}}$) / Ti (5 nm) layers
on a 128$^{\circ}$ Y-cut LiNbO$_3$ substrate, as shown in Figs.~\ref{fig1}(b) and \ref{fig1}(c),
where the parameters inside the parentheses exhibit the thickness of each layer.
To pattern the IDTs and the acoustic reflectors,
we used electron beam lithography (Elionix ELS-7700H) and deposited 35 nm of Al by electron beam evaporation for stripes of IDTs and acoustic reflectors.
Each Al stripe of an IDT has a length of 120 $\mu$m and a width of $w = 150$ nm.
Each Al stripe of the acoustic reflectors has a length of 100 $\mu$m and the same width as the IDT stripes, $w$.
All metallic stripes of the IDTs and the acoustic reflectors are separated by a distance of $d = 150$ nm.
IDT1 and IDT2 have the same structure and each of them has 20 pairs of stripes.
Each set of acoustic reflectors has 200 stripes.
After the fabrication of the acoustic cavity,
we patterned a $190 \mu\mathrm{m} \times 110 \mu\mathrm{m}$ rectangle
in between the two IDTs by photolithography (PMT D-light DL1000SG/RWC).
The Ti / CFB / Ti layers are deposited to the rectangle pattern by dc magnetron sputtering.
We fabricated samples that have the same structure but different
$t_{\mathrm{CFB}} =$ 10, 20, 25, 30, and 35 nm.

\begin{figure}[]
\centering
\includegraphics{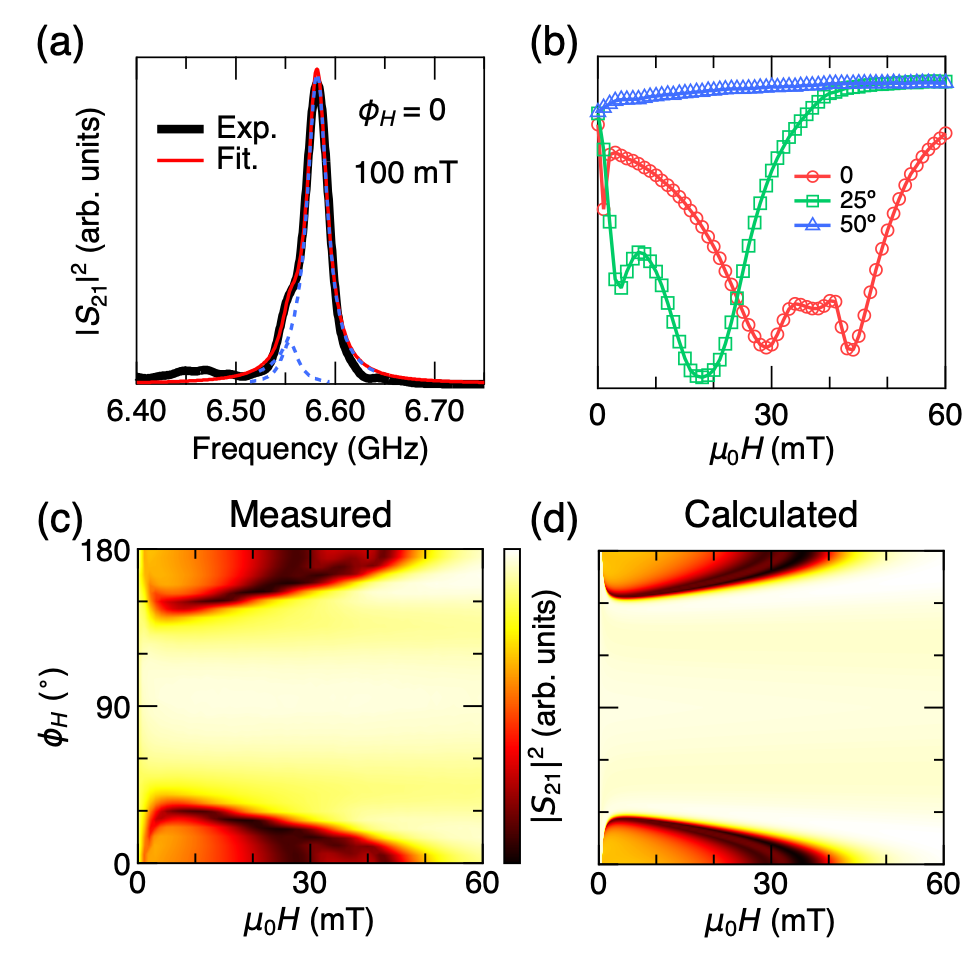}
\caption{\label{fig2} (Color online) SAW measurement results of the sample $t_{\mathrm{CFB}}$ = 20 nm.
(a) Measured SAW transmission signal ($|S_{21}|^2$) by VNA in the frequency domain.
An in-plane magnetic field of 100 mT is externally applied in a direction parallel to the SAW propagation
which is far from the resonant field of SWR driven by our SAW frequency as shown in (b), thus the transmission signal displays only the phonon response.
The black curve exhibits the measured spectrum, and the red curve shows the multiple Lorentzian fitting.
The blue dashed curves show the single Lorentzian peaks of the fitting.
(b) SAW transmission spectra at the frequency of 6.58 GHz
as a function of the amplitude of the externally applied in-plane magnetic field ($\mu_0H$).
The in-plane angles ($\phi_H$) of each curve are shown in the legend.
(c) SAW transmission spectrum at the frequency of 6.58 GHz
as a function of $\mu_0H$ and $\phi_H$.
(d) Calculated SAW transmission at the frequency of 6.58 GHz
as a function of $\mu_0H$ and $\phi_H$.}
\end{figure}

SAW transmission ($|S_{21}|^2$) is measured using a vector network analyzer (VNA)
while applying an external in-plane magnetic field $H$,
where the in-plane angle of $H$ to the SAW wavevector $\mathbf{k}$ ($\phi_H$).
This transmission is shown in Fig.~\ref{fig1}(b).
When $H$ is far enough from the resonant field of SWR driven by our SAW frequency, the SAW spectrum shows the phonon signal with no contributions from magnons, which remain unexcited.
Such a situation is shown in Fig.~\ref{fig2}(a) which depicts $|S_{21}|^2$ of the sample with $t_{\mathrm{CFB}} =$ 20 nm out of magnetic resonance when $\mu_0H = 100$ mT and $\phi_H = 0$.
The resonant frequency of the strongest SAW peak is $f_r = 6.58$ GHz.
The designed wavelength of the SAW ($\lambda_r$) is determined by the designed structure of our acoustic device;
$\lambda_r = 2(w + d) =$ 600 nm.
Using these parameters, one can calculate the SAW velocity as $v = f_r \lambda_r = 3,950$ m/s.
This value aligns with the typical SAW velocity propagating on a 128$^{\circ}$ Y-cut LiNbO\textsubscript{3}~\cite{Shibayama1976}; further supporting our assertion that this frequency peak is the resonance of the acoustic cavity.

In addition to the main resonance, we found one more peak on the lower frequency side of the main resonance [see the fitting to two Lorentzian peaks in the blue dashed curves in Fig.~\ref{fig2}(a)].
The presence of these two peaks originates from the existence of two modes allowed within our cavity device.
A detailed explanation of the origin of the two SAW peaks can be found in Sec.~1 of Ref.~\cite{SM}.
This Lorentzian fitting is used to extract the phonon linewidth $\delta_p$ to estimate the phonon relaxation rate below.

The SAW transmission out of magnetic resonance decreases when $H$ approaches the resonance field condition
since the phonons are used to excite magnons~\cite{Weiler2011,Dreher2012,Xu2018,Hwang2020,Puebla2020}.
Figure \ref{fig2}(b) shows the absorption of $|S_{21}|^2$ of the main SAW peak in Fig.~\ref{fig2}(a)
when $\phi_H = 0$, 25$^{\circ}$, and 50$^{\circ}$.
Figure \ref{fig2}(c) shows $|S_{21}|^2$ as a function of $\mu_0H$ and $\phi_H$.
In the case of a typical magnetoelastic coupling excited by a Rayleigh-type SAW,
the absorption amplitude shows maximum when $\phi_H = 45^{\circ}$\cite{Weiler2011,Dreher2012,Xu2018,Hwang2020,Xu2020,Hwang2022,Puebla2020,Yamamoto2022,Lyons2023}.
However, our results indicate that the maximum absorption across all $\phi_H$ ranges at $\phi_H \sim 25^{\circ}$ and no absorption was detected when $\phi_H > 30^{\circ}$.
This is because the magnon dispersion is raised to a higher frequency due to the dipolar field and does not meet the phonon dispersion at $\phi_H > 30^{\circ}$ (Fig.~7 of Ref.~\cite{SM}).
Note that the in-plane uniaxial magnetic anisotropy of our CFB film~\cite{Ito2022} aligned in-plane perpendicular to $\mathbf{k}$ causes the dips of SAW transmission around $\mu_0H = 0$.
Assuming the uniaxial magnetic anisotropy and the magnon–phonon coupling strength obtained by experiments (see below) allow us to calculate the SAW transmission as a function of an external in-plane magnetic field.
As shown in Fig.~\ref{fig2}(d), the calculation agrees well with the measured result.
A detailed description of the calculation can be found in Sec.~5A of Ref.~\cite{SM}.

The distinct observation of anticrossing, represented by split features in SAW absorption, becomes most pronounced when $\phi_H \sim 0$.
This splitting does not originate from the longitudinal strain, commonly considered the dominant one in SAWs generated on a 128$^{\circ}$ Y-cut LiNbO$_3$ substrate.
The strain tensor is expressed as $\varepsilon_{ij} = (\partial_j u_i + \partial_i u_j)/2$,
where $u_i$ and $u_j$ ($i,j = x,y,z$) are components of the elastic deformation vector field.
When the longitudinal component $\varepsilon_{xx}$ couples to magnons,
the maximum magnetoelastic coupling occurs when the angle between the SAW propagation and the in-plane magnetization $\phi$ is 45$^{\circ}$,
whereas the shear component $\varepsilon_{xy}$ of the SAW shows the maximum magnetoelastic coupling
when $\phi =$ 0 or 90$^{\circ}$~\cite{Dreher2012}.
While in our device, $\varepsilon_{xx}$ is larger than $\varepsilon_{xy}$ (Fig.~2(c) of Ref.~\cite{SM}),
$\varepsilon_{xx}$ decreases abruptly away from the surface due to a change of sign of $u_x$~\cite{Babu2021}.
Therefore, $\varepsilon_{xx}$ faces a limitation in terms of penetration depth,
which results in an insufficient coupling strength to observe magnon–phonon anticrossing.
On the contrary, $\varepsilon_{xy}$ has a larger penetration depth, making the associated coupling strength dominant.
Furthermore, the observation of significant split features is strongly supported by the fact that when the phonon (magnon) wavevector aligns parallel to the external magnetic field, the transverse strain can exhibit substantial magnetoelastic coupling, whereas the longitudinal strain cannot~\cite{Hioki2022,Rezende}.

\begin{figure}[]
\centering
\includegraphics{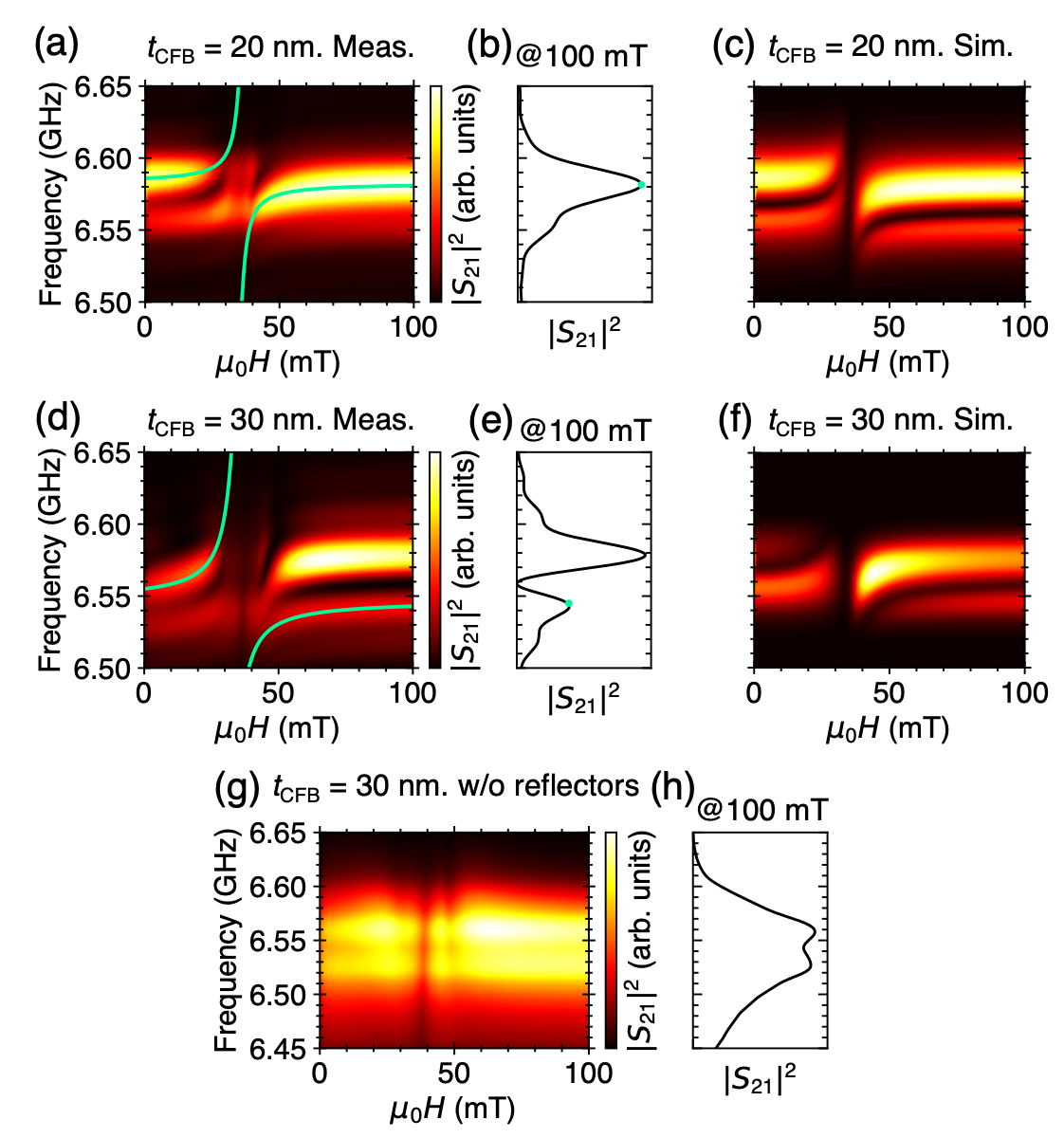}
\caption{\label{fig3} (Color online) SAW transmissions when the external magnetic field is applied in the direction of SAW propagation; $\phi_H = 0$.
(a),(b) SAW transmission signal ($|S_{21}|^2$) of the sample with $t_{\mathrm{CFB}} = 20$ nm under (a) various amplitudes of the magnetic field $\mu_0H$ and (b) $\mu_0H =$ 100 mT.
The green marker in (b) represents the local maximum used for the anticrossing fitting,
shown as the green curves in (a).
(c) Calculated SAW transmission of the sample with $t_{\mathrm{CFB}} = 20$ nm
as a function of the frequency and $\mu_0H$.
(d),(e) $|S_{21}|^2$ of the sample with $t_{\mathrm{CFB}} = 30$ nm
under (d) various $\mu_0H$ and (e) $\mu_0H =$ 100 mT.
The green marker in (e) represents the local maximum used for the anticrossing fitting, shown as the green curves in (d).
(f) Calculated SAW transmission of the sample with $t_{\mathrm{CFB}} = 30$ nm
as a function of the frequency and $\mu_0H$.
(g),(h) $|S_{21}|^2$ of the sample with $t_{\mathrm{CFB}} = 30$ nm,
but in the absence of acoustic reflectors (g) under various $\mu_0H$
and (h) $\mu_0H =$ 100 mT.}
\end{figure}

Having clarified the origin of the coupling, we now focus on the magnon–phonon coupling at $\phi_H = 0$,
where the magnetoelastic coupling is dominated by the shear strain $\varepsilon_{xy}$.
Figure \ref{fig3}(a) shows the SAW transmission signal $|S_{21}|^2$ of the sample
with $t_{\mathrm{CFB}} = 20$ nm as a function of frequency and $\mu_0 H$,
and Fig.~\ref{fig3}(b) the SAW transmission spectrum when $\mu_0 H = 100$ mT.
As we mentioned above, a SAW peak exists at a lower frequency side to the main peak, giving rise to multiple anticrossing features.
However, as each SAW mode couples only to the magnon mode with its same wavenumber, each SAW branch shows a single anticrossing.
Therefore, we focus on anticrossing of one mode per device to estimate its coupling.
We first fit the phonon branches in $\omega$--$H$ dispersion
taken by the local maximum of the SAW spectrum [see the marker in Fig.~\ref{fig3}(b)]
at each field with our magnon–phonon coupling model:
\begin{equation}\label{eq1}
    \omega^2 = \frac{\omega_m^2 + \omega_p^2}{2} \pm \frac{1}{2}
    \sqrt{\left(\omega_m^2 - \omega_p^2 \right)^2 + \left(2\delta\omega_{\mathrm{bare}}^2 \right)^2},
\end{equation}
where $\omega_m$ and $\omega_p$ are the magnon and phonon resonant frequencies.
The derivation of Eq.~(\ref{eq1}) and the definition of $\delta\omega_{\mathrm{bare}}$ can be found in Sec.~4 of Ref.~\cite{SM}.
The anticrossing fitting is shown in Fig.~\ref{fig3}(a) as green curves.

From the parameters obtained by the fitting, we reproduce the anticrossing detected by SAW transmission spectra using our SAW transmission model.
We modeled two phonon modes coupling to each magnon mode corresponding to its wavenumber. 
For the details of the model, see Sec.~5 of Ref.~\cite{SM}.
As a result, the SAW transmission is well reproduced as shown in Fig.~\ref{fig3}(c).

Furthermore, in Figs.~\ref{fig3}(d)--\ref{fig3}(f), we present the same analysis as shown in Figs.~\ref{fig3}(a)--\ref{fig3}(c)
but for the sample with $t_{\mathrm{CFB}} = 30$ nm.
It is notable that the upper phonon branch of the main SAW peak in Fig.~\ref{fig3}(d) is vaguely visible at $\mu_0H \sim 0$,
however, it is no longer detectable at $0 < \mu_0 H < 40$ mT,
as this peak shifts out of the frequency range of our cavity due to a redshift given by the strong mode interaction.
Therefore, we fitted anticrossing of the peak at 6.54 GHz in Fig.~\ref{fig3}(e) to obtain the parameters as shown in Fig.~\ref{fig3}(d).
The results and calculations of samples with other $t_{\mathrm{CFB}}$ are shown in Fig.~8 of Ref.~\cite{SM}.

Additionally, it should be noted that a device with the same structure as used in this experiment
but without the presence of the acoustic reflectors does not show any anticrossing as shown in Figs.~\ref{fig3}(g) and \ref{fig3}(h). This is due to the considerably higher phonon relaxation and smaller coupling strength originating from smaller $\varepsilon_{xy}$ when there is an absence of reflectors that form an acoustic cavity (Secs.~2 and 3B of Ref.~\cite{SM}).

\begin{figure}[]
\centering
\includegraphics{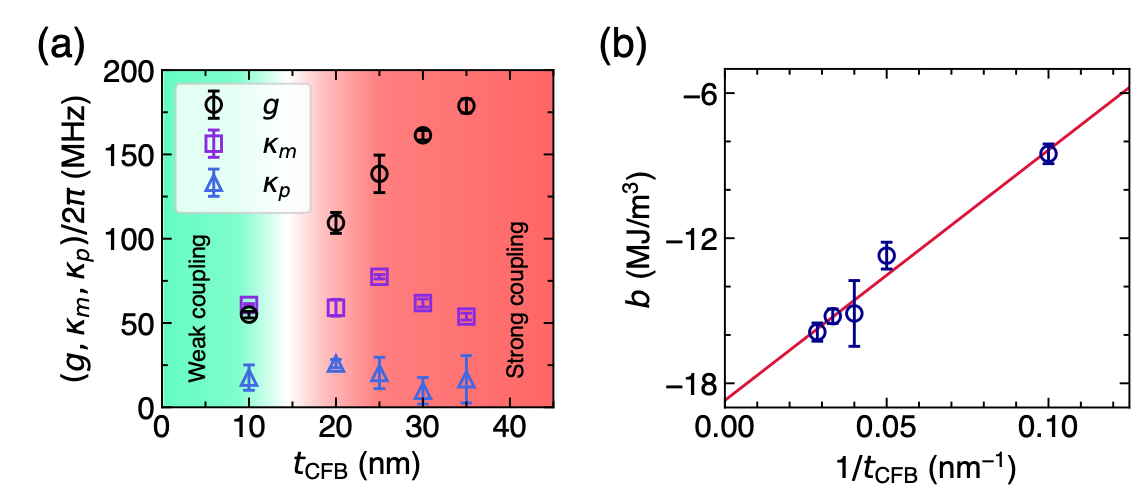}
\caption{\label{fig4} (Color online)
(a) The coupling strength ($g$; the black curve) and magnon ($\kappa_m$; the purple square) and phonon ($\kappa_p$; the blue triangle) relaxation rates as a function of the thickness of the CFB layer.
(b) The effective magnetoelastic constant as a function of the inverse of the thickness of the CFB layer.
The solid line exhibits the surface magnetoelastic coupling fitting.
The data points and error bars of (a) and (b) are the mean and s.d. of measurements of three or more different devices, respectively.}
\end{figure}

Lastly, we present the magnon–phonon coupling estimation in the devices with varying CFB thicknesses.
Figure \ref{fig4}(a) shows the coupling strength $g$
as a function of $t_{\mathrm{CFB}}$ taken by the anticrossing fittings in Fig.~\ref{fig3} and Fig. 8 of Ref.~\cite{SM} with Eq.~(\ref{eq1}) and Eqs.~(9), (12) of Ref.~\cite{SM}.
While our magnon–phonon coupling model (Sec.~4 of Ref.~\cite{SM}) predicts $g \sim \sqrt{t_{\mathrm{CFB}}}$,
this prediction does not hold true, as shown in Fig.~\ref{fig4}(a), due to variations in the effective magnetoelastic coupling coefficient $b$ with changes in the thickness of the ferromagnetic layer $t_{\mathrm{FM}}$.
It is known that $b$ is determined by contributions of the bulk ($b_v$) and surface ($b_s$) magnetoelastic couplings~\cite{Sun1991,Chumak2021,Gowtham2016} as $b = b_v + b_s/t_{\mathrm{CFB}}$.
Figure \ref{fig4}(b) shows $b$ as a function of the inverse of $t_{\mathrm{CFB}}$, determined by the values of $g$ and Eq.~(12) of Ref.~\cite{SM}.
The fitting of $b$ with the surface magnetoelastic coupling shown as a solid line yields
$b_v = -18.7$ MJ/m$^3$ and $b_s = 104$ mJ/m$^2$, where the values are in agreement with the known value for CFB~\cite{Gowtham2016}.

Next, to evaluate the attainment of strong coupling we estimate the relaxation rates of magnons ($\kappa_m$) and phonons ($\kappa_p$).
To obtain $\kappa_m$, we used the Gilbert damping $\alpha$ of our CFB films as $\kappa_m = \omega_m \alpha$.
The Gilbert damping is measured by ferromagnetic resonance (FMR) using coplanar waveguides (see Sec.~6 and Fig.~9 of Ref.~\cite{SM}).
For $\kappa_p$, we obtain the linewidth of $|S_{21}|^2$ peaks ($\delta_p$)
by Lorentzian fittings as shown as the dashed curves in Fig.~\ref{fig2}(a) and determine $\kappa_p = 2\pi\delta_p$.
The estimated relaxation rates of devices with varying $t_{\mathrm{CFB}}$ are depicted alongside the coupling strength in Fig. \ref{fig4}(a). 
As one can find from Fig. \ref{fig4}(a), $\kappa_m$ is always larger than $\kappa_p$ in our devices, thus confirming the ratio $\Gamma = g / \kappa_m > 1$ attains strong coupling. For the devices with $t_{\mathrm{CFB}} \geq 20$ nm, $\Gamma > 1$, i.e., strong coupling is achieved. 
However, for the devices with $t_{\mathrm{CFB}} = 10$ nm, $\Gamma < 1$, thus it is not in the strong coupling regime.
We would also like to note that, the FMR experiments with the same Ti/CFB/Ti layer using photon excitations by coplanar waveguides do not show evident anticrossing features (see Fig. 9(a) of Ref.~\cite{SM}).
Therefore, the anticrossing behavior presented in this work seems to be exclusive to strong magnon–phonon coupling enabling the potential existence of a magnon–phonon quasiparticle called the magnon–polaron.

To summarize, we achieved strong coupling between magnons and SAW phonons in Co\textsubscript{20}Fe\textsubscript{60}B\textsubscript{20} thin films using an acoustic cavity constructed by a two-port SAW resonator.
SAW phonon anticrossings are observed when an external magnetic field is applied parallel to the SAW propagation direction.
By fitting anticrossings with our theoretical model, we estimated the coupling strength of our devices.
For the devices with $t_{\mathrm{CFB}} \geq$ 20 nm, we confirmed the achievement of strong coupling.
The variety in selecting magnetic materials and the usage of well-established SAW devices, facilitated by the independent behavior of magnons and phonons in our system, will pave the way to explore magnon–phonon strong coupling physics with on-chip devices at room temperature.
Moreover, it holds the potential to offer insights into studying coherently coupled magnon–phonon hybridized quasiparticles enabling the development of magnetic field-controlled acoustic devices and less lossy magnon-based information processing devices.
Besides, the ascending coupling strength achieved by increasing the thickness of the magnetic material within our model, combined with the flexibility in selecting magnetic materials, makes provision for achieving magnon–phonon ultrastrong coupling where the ratio between the coupling strength and the resonant frequency is comparable; $g / \omega_{\mathrm{res}} \gtrsim 0.1$~\cite{Kockum2019}.
According to our magnon–phonon coupling model, employing a material with a 4 times higher magnetoelastic coupling coefficient than CFB and a saturation magnetization of 1 T, such as Tb\textsubscript{0.3}Dy\textsubscript{0.7}Fe\textsubscript{2}~\cite{Sandlund1994}, with the same device design we used, and the material thickness of 20 nm, one could potentially reach $g / \omega_{\mathrm{res}} \sim 0.1$.

\begin{acknowledgments}
This work was supported by Grants-in-Aid for Scientific Research (S) (No. 19H05629) and the Japan Society for the Promotion of Science Grants-in-Aid for Scientific Research (No. 20H01865). Y.H. thanks to RIKEN Junior Research Associate Program for supporting this work. L.L. would like to thank the support from JSPS through Research Program for Young Scientists (No. 23KJ0778). F.C. and W.L. acknowledge China National Key Research and Development Plan (2022YFE0103300), and Y.O. acknowledges the RIKEN-China cooperation project. The authors thank Kei Yamamoto for fruitful discussions.
\end{acknowledgments}

\bibliography{01_ms}

\end{document}


\preprint{Supplemental Material}

\title{{\Large Supplemental Material}\\
Strongly Coupled Spin Waves and Surface Acoustic Waves at Room Temperature}

\author{Yunyoung Hwang}
\affiliation{%
 Institute for Solid State Physics, University of Tokyo, Kashiwa 277-8581, Japan
}%
\affiliation{%
 CEMS, RIKEN, 2-1, Hirosawa, Wako 351-0198, Japan
}%
\author{Jorge Puebla}%
 \email{jorgeluis.pueblanunez@riken.jp}
\affiliation{%
 CEMS, RIKEN, 2-1, Hirosawa, Wako 351-0198, Japan
}%


\author{Kouta Kondou}
\affiliation{%
 CEMS, RIKEN, 2-1, Hirosawa, Wako 351-0198, Japan
}%

\author{Carlos Gonzalez-Ballestero}
\affiliation{%
 Institute for Quantum Optics and Quantum Information of the Austrian Academy of Sciences, A-6020 Innsbruck, Austria
}%
\affiliation{%
 Institute for Theoretical Physics, University of Innsbruck, A-6020 Innsbruck, Austria
}%

\author{Hironari Isshiki}
\affiliation{%
 Institute for Solid State Physics, University of Tokyo, Kashiwa 277-8581, Japan
}%

\author{Carlos S\'{a}nchez Mu\~{n}oz}
\affiliation{%
 Departamento de F\'{i}sica Te\'{o}rica de la Materia Condensada and Condensed Matter Physics Center (IFIMAC), Universidad Aut\'{o}noma de Madrid, 28049 Madrid, Spain
}%

\author{Liyang Liao}
\affiliation{%
 Institute for Solid State Physics, University of Tokyo, Kashiwa 277-8581, Japan
}%

\author{Fa Chen}
\affiliation{%
 School of Integrated Circuits, Wuhan National Laboratory for
Optoelectronics, Huazhong University of Science and Technology, Wuhan 430074,
People’s Republic of China.
}%

\author{Wei Luo}
\affiliation{%
 School of Integrated Circuits, Wuhan National Laboratory for
Optoelectronics, Huazhong University of Science and Technology, Wuhan 430074,
People’s Republic of China.
}%

\author{Sadamichi Maekawa}
\affiliation{%
 CEMS, RIKEN, 2-1, Hirosawa, Wako 351-0198, Japan
}%
\affiliation{%
 Advanced Science Research Center, Japan Atomic Energy Agency, Tokai 319-1195, Japan
}%
\affiliation{%
 Kavli Institute for Theoretical Sciences, University of Chinese Academy of Sciences, Beijing 100049, People’s Republic of China
}%

\author{Yoshichika Otani}
\email{yotani@issp.u-tokyo.ac.jp}
\affiliation{%
 Institute for Solid State Physics, University of Tokyo, Kashiwa 277-8581, Japan
}%
\affiliation{%
 CEMS, RIKEN, 2-1, Hirosawa, Wako 351-0198, Japan
}%


                              
\setcounter{section}{0}
\setcounter{equation}{0}
\setcounter{figure}{0}
\setcounter{table}{0}
\setcounter{page}{1}
\renewcommand{\thesection}{S\arabic{section}}
\renewcommand{\theequation}{S\arabic{equation}}
\renewcommand{\thefigure}{S\arabic{figure}}
\renewcommand{\thetable}{S\arabic{table}}
\renewcommand{\bibnumfmt}[1]{[S#1]}
\renewcommand{\citenumfont}[1]{S#1}

\maketitle
\tableofcontents

\section{SAW modes in our acoustic cavity devices}
A distinctive feature of an optical or acoustic cavity is the emergence of a set of peaks separated by the same frequency spacing representing the set of standing wave modes inside the cavity. This particular frequency spacing is known as free spectral range (FSR) and can be estimated by $v/(2L_c)$, where $v$ is the wave velocity and $L_c$ is the cavity length. And, it has been observed in two-port SAW resonator devices~\cite{Hatanaka2022,Hwang2023}. For the case of the acoustic cavity devices of the present study, if we assume the cavity length as $L_c \geq L$, where $L$ is the length of propagation, we estimate a separation by the FSR of $\leq 10$ MHz. However, in our transmission/absorption measurements we hardly observed this set of peaks with a separation of $\leq 10$ MHz. We understand it as consequence of the linewidth broadening due to the mass loading of Ti/CFB/Ti layers. 

\begin{figure}[h]
\centering
\includegraphics[width=0.45\textwidth]{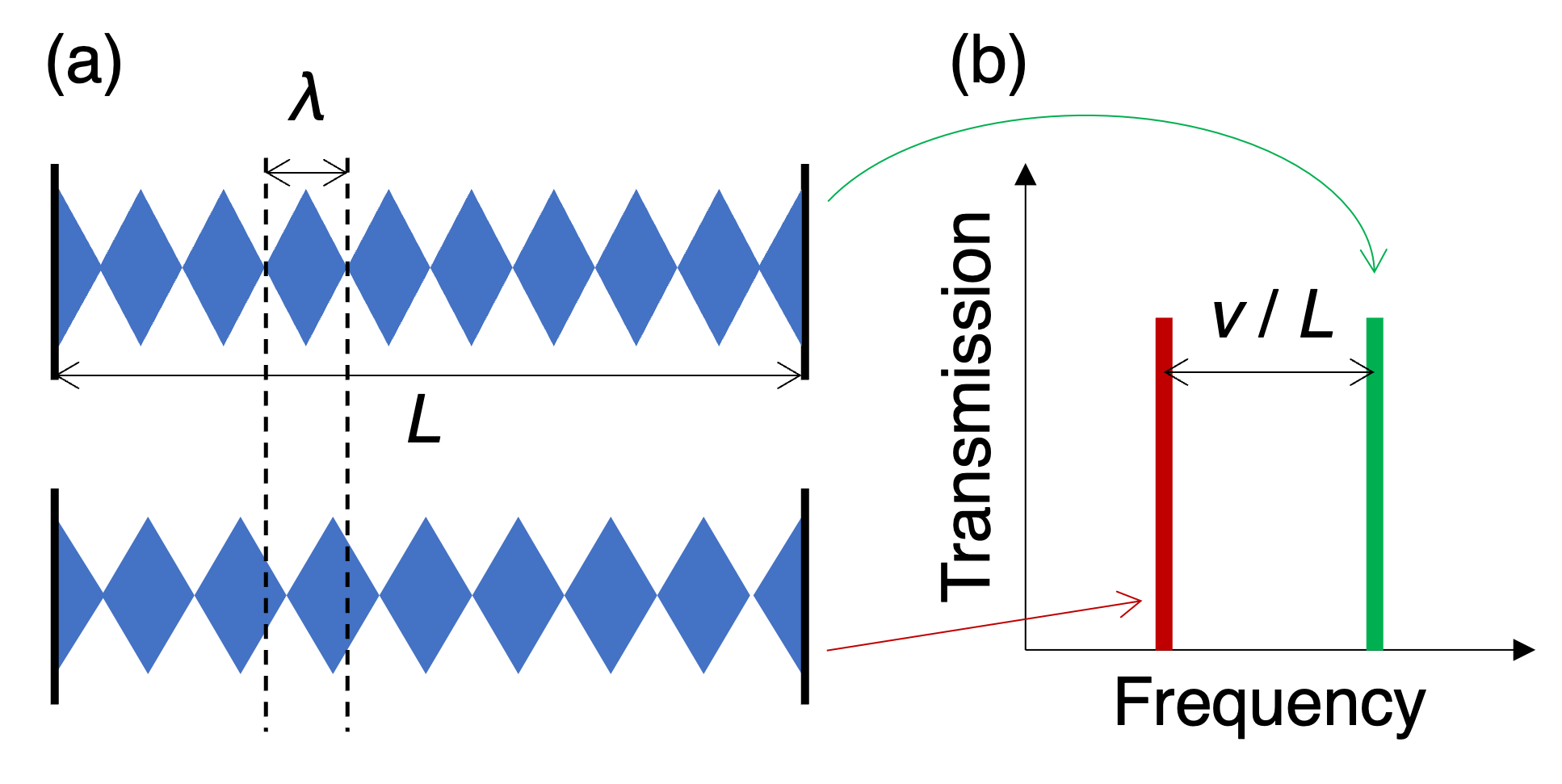}
\caption{\label{fig:phase} SAW phase shift due to the propagation.
(a) Schematic illustration of SAWs with two different wavelengths in the same propagation length.
(b) Schematic illustration of separated SAW peaks due to the phase shift in a frequency domain.}
\end{figure}

Instead, in our transmission/absorption measurements we clearly observe two modes within our cavity [see Fig.~2(a) in the main text]. We understand it as follows: i) first, needless to say in an ideal device we should expect a single well-defined peak representing a single mode, however, the fabrication resolution sets a limit, in our case, given by the resolution of the electron beam lithography system used to fabricate our acoustic wave device. This limit of fabrication resolution is shown as the linewidth broadening of our transmission/absorption measurements, which is approximately equivalent to fluctuations of IDTs dimensions of $\sim$ 8 nm. ii) Second, as the nominal distance between our IDT sets or propagation length is a multiple of $\lambda_r$, the existence of additional modes requires to fulfill the phase shift condition given by $e^{ikL} = e^{i2\pi L/\lambda_r} = 1$, defining the arrival frequency of an additional mode to the detecting IDT set. The frequency intervals can be simply related to a period of $v/L$, as shown in Fig.~\ref{fig:phase}, where $v$ is the velocity of the SAW, and $L=$ 210 $\mu$m is not the cavity length but the propagation length from the generating IDT set to the detecting IDT set, resulting in approximately $\sim$ 19 MHz, in proximity to the frequency separation we observe in our transmission/absorption measurements.     

\section{SAW strain simulations}\label{sec:strain}

\begin{figure}
\centering
\includegraphics[width=0.8\textwidth]{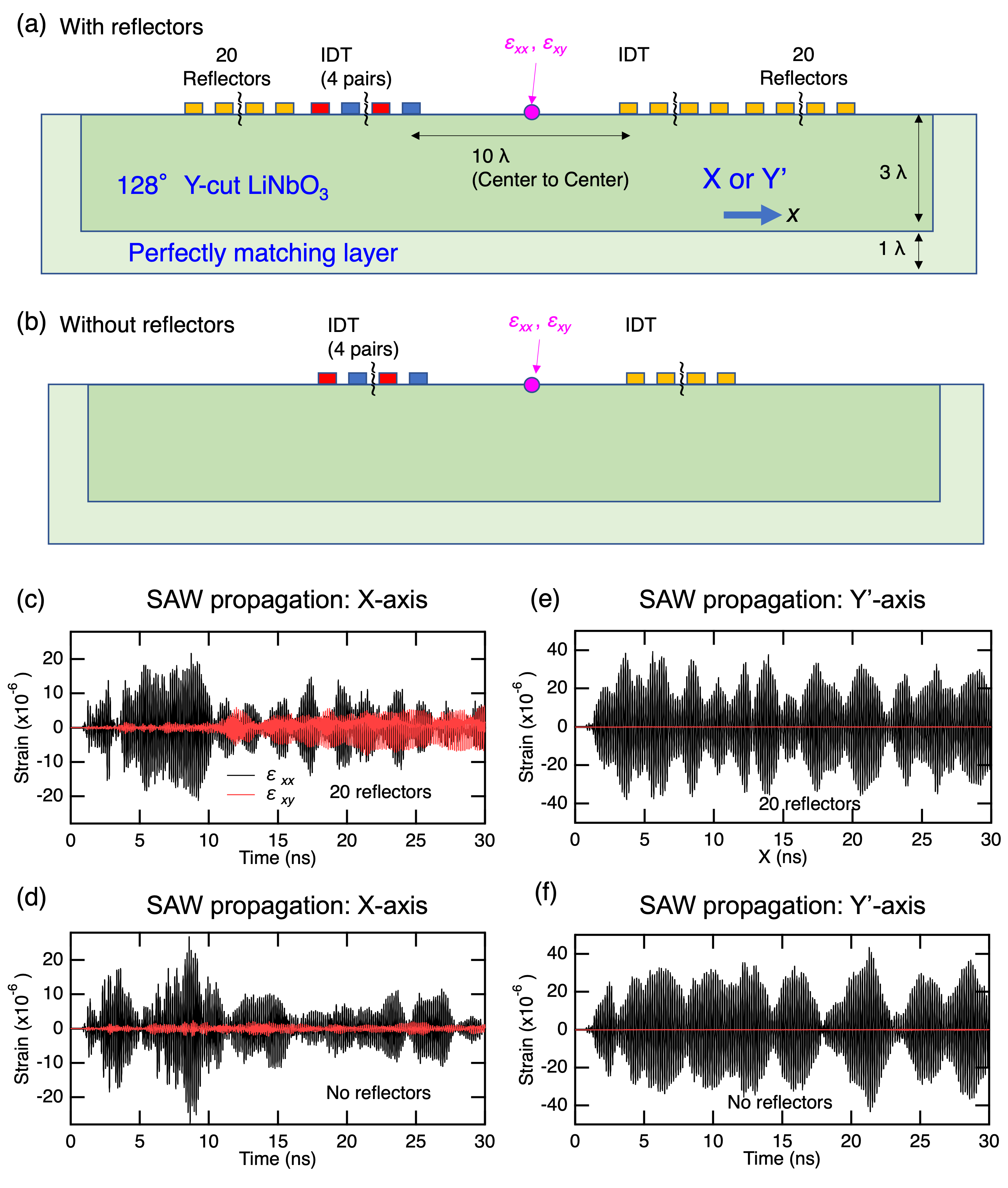}
\caption{\label{fig:COMSOL} Simulation of the SAW strain components.
(a),(b) Schematic description of the geometry of the simulation (a) with and (b) without reflectors.
(c),(d) Simulated strain components as a function of time (a) with and (b) without reflectors with SAWs propagating along the X-axis.
(e),(f) Same as (c) and (d), but with SAWs propagating along the Y'-axis.}
\end{figure}

We performed simulations to confirm the existence of SAW strain components
$\varepsilon_{xx}$ and $\varepsilon_{xy}$ using COMSOL Multiphysics\textsuperscript{\textregistered}.
For simplicity, 4 pairs of signal and ground lines of an IDT set are simulated.
We have set two geometries:
one with two sets of 20 reflectors [Fig.~\ref{fig:COMSOL}(a)];
and the other without reflectors [Fig.~\ref{fig:COMSOL}(b)].

The widths of one stripe of IDTs and reflectors, and the gap between them are 150 nm.
In accordance with the IDT design, we set the SAW wavelength $\lambda = 600$ nm.
We simulate SAWs propagating on the X- and Y'-axes of the 128$^{\circ}$ Y-cut LiNbO$_3$.
We set the SAW velocity $v = 4,000$ m/s for the SAW propagating along the X-axis,
and $v = 3,665$ m/s for the SAW propagating along the Y'-axis,
which have been calculated independently by the eigenfrequency analyses.
The distance between the two IDT sets is $10\lambda$ and the thickness of the substrate is $3\lambda$.
We set a $\lambda$-thick perfectly matching layer at the bottom of the substrate.
The mesh size for the simulation is $\sim \lambda/8$.
The input frequency for the IDT is $v / \lambda$.

Figures \ref{fig:COMSOL}(c) and \ref{fig:COMSOL}(d) show the strain components with and without reflectors, respectively, with the SAW propagating along the X-axis.
Figures \ref{fig:COMSOL}(e) and \ref{fig:COMSOL}(f) show the same simulation as Figs.~\ref{fig:COMSOL}(c) and \ref{fig:COMSOL}(d)
but with the SAW propagating along the Y'-axis.
It is important to note that, we defined the $x$-axis as always parallel to the SAW propagation direction, thus $\varepsilon_{xx}$ is always the longitudinal strain of SAWs in Figs.~\ref{fig:COMSOL}(c)--\ref{fig:COMSOL}(f).
From the simulation results, one can confirm that employing an acoustic cavity structure significantly enhances $\varepsilon_{xy}$ when the SAW propagates parallel to the X-axis.

To better account for the strain components, we calculate the root mean square (RMS) of each strain component as
$\varepsilon_{xx}^{\mathrm{RMS}}$ and $\varepsilon_{xy}^{\mathrm{RMS}}$ in a time range 25--30 ns
and calculate the ratio $\eta = \varepsilon_{xy}^{\mathrm{RMS}} / \varepsilon_{xx}^{\mathrm{RMS}}$.
As a result, we obtain $\eta = 0.95$ for the simulation with 20 reflectors [Fig.~\ref{fig:COMSOL}(c)]
and $\eta = 0.15$ for the simulation without reflectors [Fig.~\ref{fig:COMSOL}(d)].
We subsequently employ $\eta = 0.5$ as a compromise since our simulations have been done
using a simplified geometry in comparison to our actual device design.



\section{Additional measurements to confirm contributions of the strain components}
\subsection{Different propagation direction of SAWs}
In the above, in Figs.~\ref{fig:COMSOL}(c)--\ref{fig:COMSOL}(f),
we confirmed that the dominant SAW strain is the longitudinal strain $\varepsilon_{xx}$,
however, $\varepsilon_{xy}$ is significant for SAWs propagating along the crystal X-axis of a 128$^{\circ}$ Y-cut LiNbO$_3$ substrate, whereas $\varepsilon_{xy}$ is negligible compare to $\varepsilon_{xx}$ when SAWs propagate along the Y'-axis, which is in-plane perpendicular to the X-axis.

\begin{figure}
\centering
\includegraphics[width=0.5\textwidth]{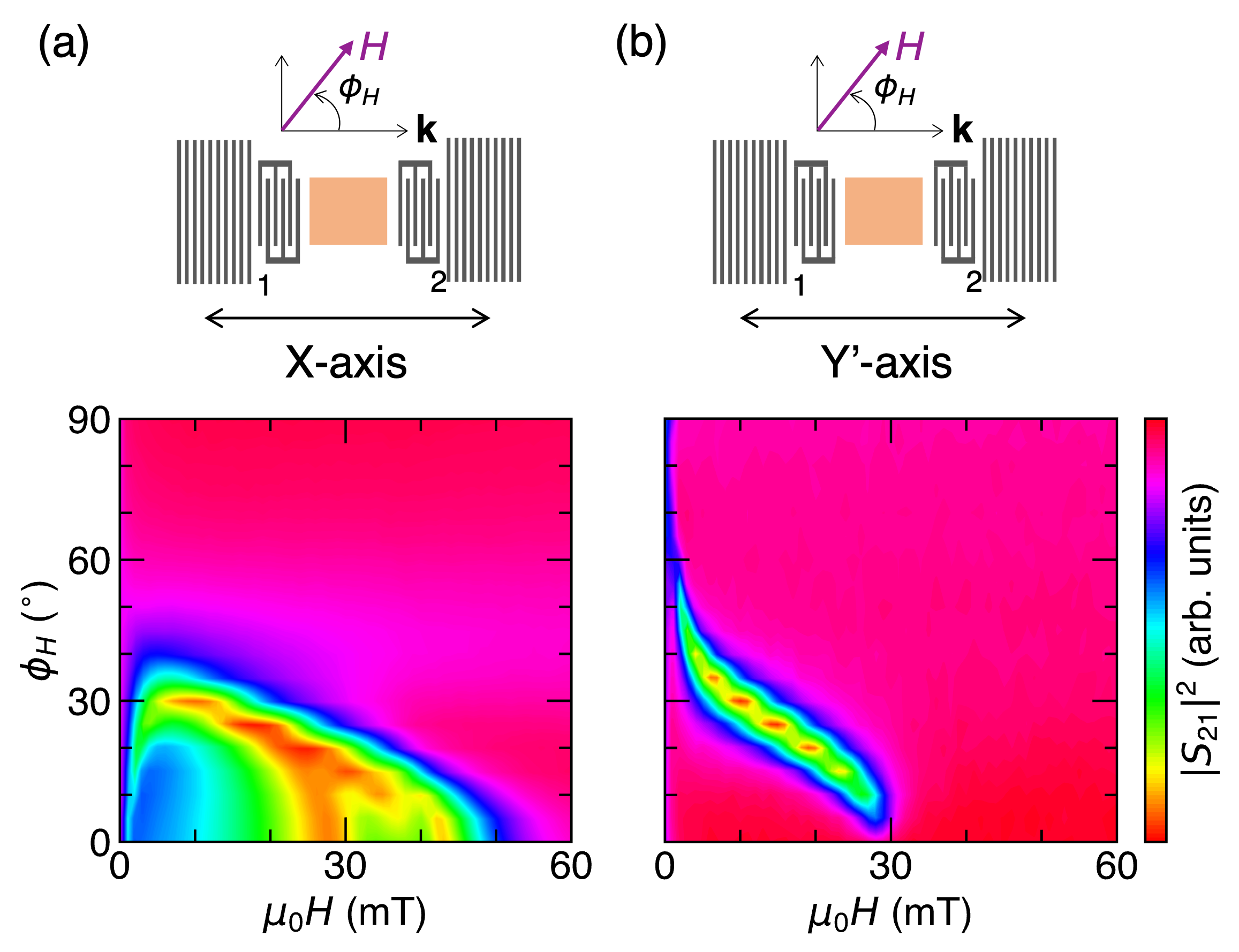}
\caption{\label{fig:XandY} In-plane magnetic field angle dependence of
SAW transmissions propagating on the X- and Y'-axes of the samples with $t_{\mathrm{CFB}} = 20$ nm.
(a),(b) SAW transmission spectra at the frequency of 6.58 GHz
as a function of the amplitude and in-plane angle $\phi_H$ of the applied magnetic field $\mu_0 H$ 
when $\mathbf{k}$ propagates along (a) the X-axis and (b) Y'-axis of the LiNbO$_3$ substrate.}
\end{figure}

To confirm the magnetoelastic coupling by $\varepsilon_{xy}$ occurred in our device,
we fabricated the same-structured devices as the one used for the results in the main text
with the different propagation direction; the Y'-axis.
The SAW transmissions at the resonant frequency at various in-plane angles to $\mathbf{k}$
of the external magnetic field are shown in Fig.~\ref{fig:XandY}.
Note that since the difference in the SAW velocity for different propagation directions,
the resonant frequencies are 6.58 GHz and 6.06 GHz for the devices
of which the SAW propagates along the X- and Y'-axis, respectively.
On the one hand, when SAWs propagate along the X-axis [Fig.~\ref{fig:XandY}(a)],
as the result shown in the main text,
one can find significant SAW absorptions at $\phi_H \sim 0$.
On the other hand, when SAWs propagate along the Y'-axis [Fig.~\ref{fig:XandY}(b)],
SAW absorptions are not observed at $\phi_H \sim 0$,
which implies the lack of $\varepsilon_{xy}$ results in the lack of the magnetoelastic coupling
at $\phi_H \sim 0$.
Moreover, from the device of which the SAW propagates along the Y'-axis [Fig.~\ref{fig:XandY}(b)],
we did not observe anticrossing at any $\phi_H$.

\subsection{Magnon--phonon coupling of devices with thin CFB}
As shown in Ref.~\cite{Babu2021}, the magnetoelastic coupling by $\varepsilon_{xx}$ is substantially suppressed even though the local value of $\varepsilon_{xx}$ is significant
due to its rapid change in a thickness direction of a ferromagnetic film.
In contrast, $\varepsilon_{xy}$ does not decrease rapidly in the thickness direction,
thus the thicker ferromagnet layer allows for more magnon--phonon interactions by $\varepsilon_{xy}$,
leading to a higher effective coupling strength.
To confirm this, we fabricated the same-structured device as the main text but with a thinner CFB; $t_{\mathrm{CFB}} = 3$ nm.

\begin{figure}
\centering
\includegraphics[width=0.5\textwidth]{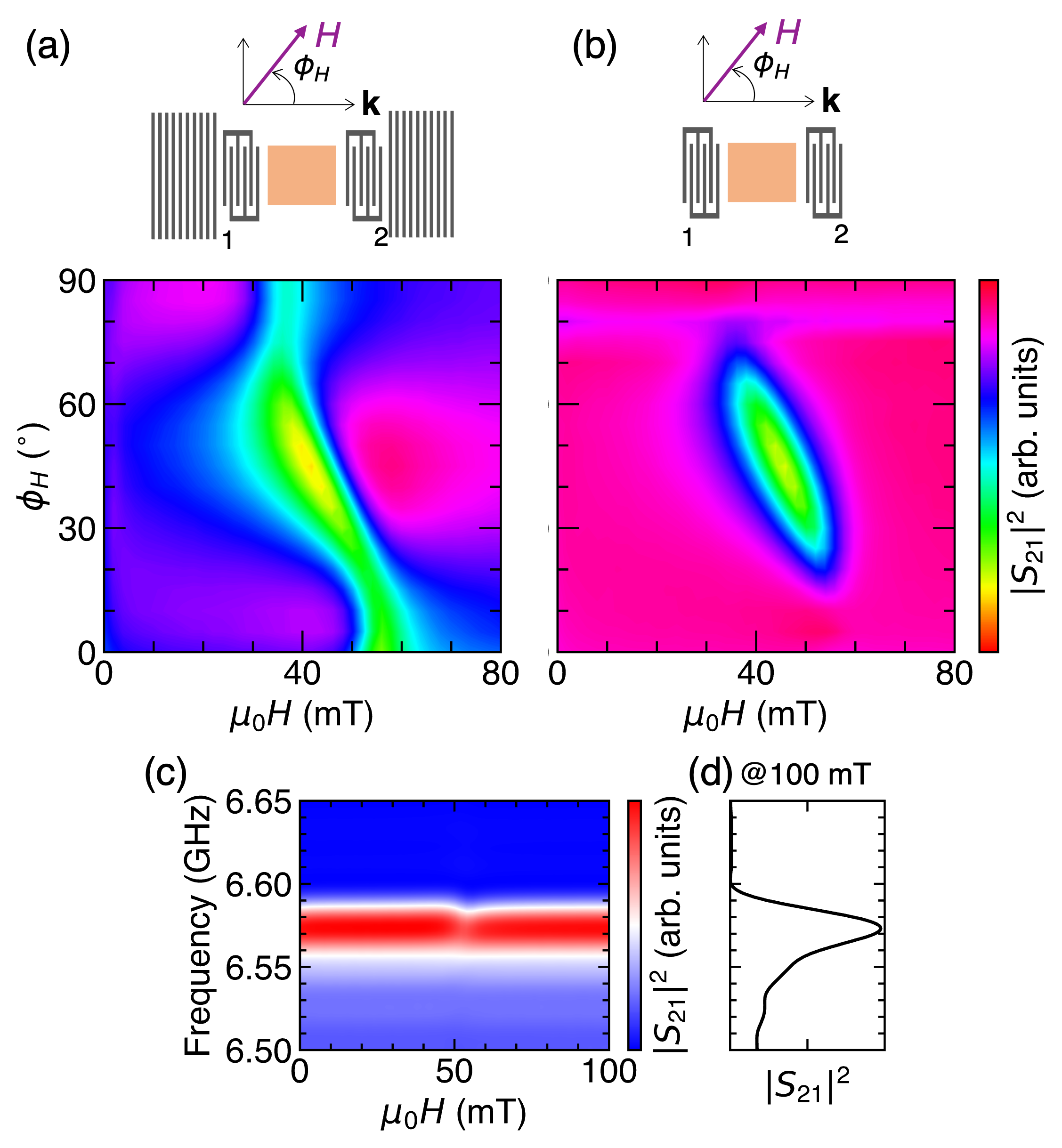}
\caption{\label{fig:thin} SAW transmission measurements of the sample
with $t_{\mathrm{CFB}} = 3$ nm.
(a),(b) SAW transmission spectrum at the frequency of 6.58 GHz
as a function of the amplitude and in-plane angle $\phi_H$ of the applied magnetic field $\mu_0 H$ 
of the samples (a) with reflectors and (b) without reflectors.
(c),(d) SAW transmission signal of the sample with reflectors
under (c) various amplitudes of the magnetic field and (d) 100 mT of the magnetic field.
The SAWs in (a) and (b) propagate along the X-axis.}
\end{figure}

The SAW absorptions at the frequency of 6.58 GHz at various $\phi_H$
of a device including reflectors and a device in the absence of reflectors
are shown in Fig.~\ref{fig:thin}(a) and \ref{fig:thin}b, respectively.
As shown in Sec.~\ref{sec:strain}, $\varepsilon_{xy}$ is enhanced by reflectors,
thus only the device with reflectors shows significant SAW absorptions at $\phi_H \sim 0$.
We note that the result in Fig.~\ref{fig:thin}(b) shows a similar behavior
as our previous result of 1.6-nm-thick CFB~\cite{Xu2020}.

In addition, due to the lower saturation magnetization of the thin CFB than the devices
with $t_{\mathrm{CFB}} \geq 10$ nm~\cite{Tarequzzaman2018}, magnon--phonon coupling can occur at $\phi_H = 90^{\circ}$
due to the lower dipolar field [see Eqs. (\ref{eq:Hz}--\ref{eq:Hphi}), Sec.~\ref{sec:angDep}, and Fig.~\ref{fig:SWR_angle}],
thus one observes SAW absorption.
Therefore, we confirm the typical in-plane angular dependence of the magnetoelastic coupling
by $\varepsilon_{xy}$, which is proportional to $\cos{2\phi}$ (see the next section).

\section{Magnon--phonon coupling model}
\begin{figure}
\centering
\includegraphics[width=0.5\textwidth]{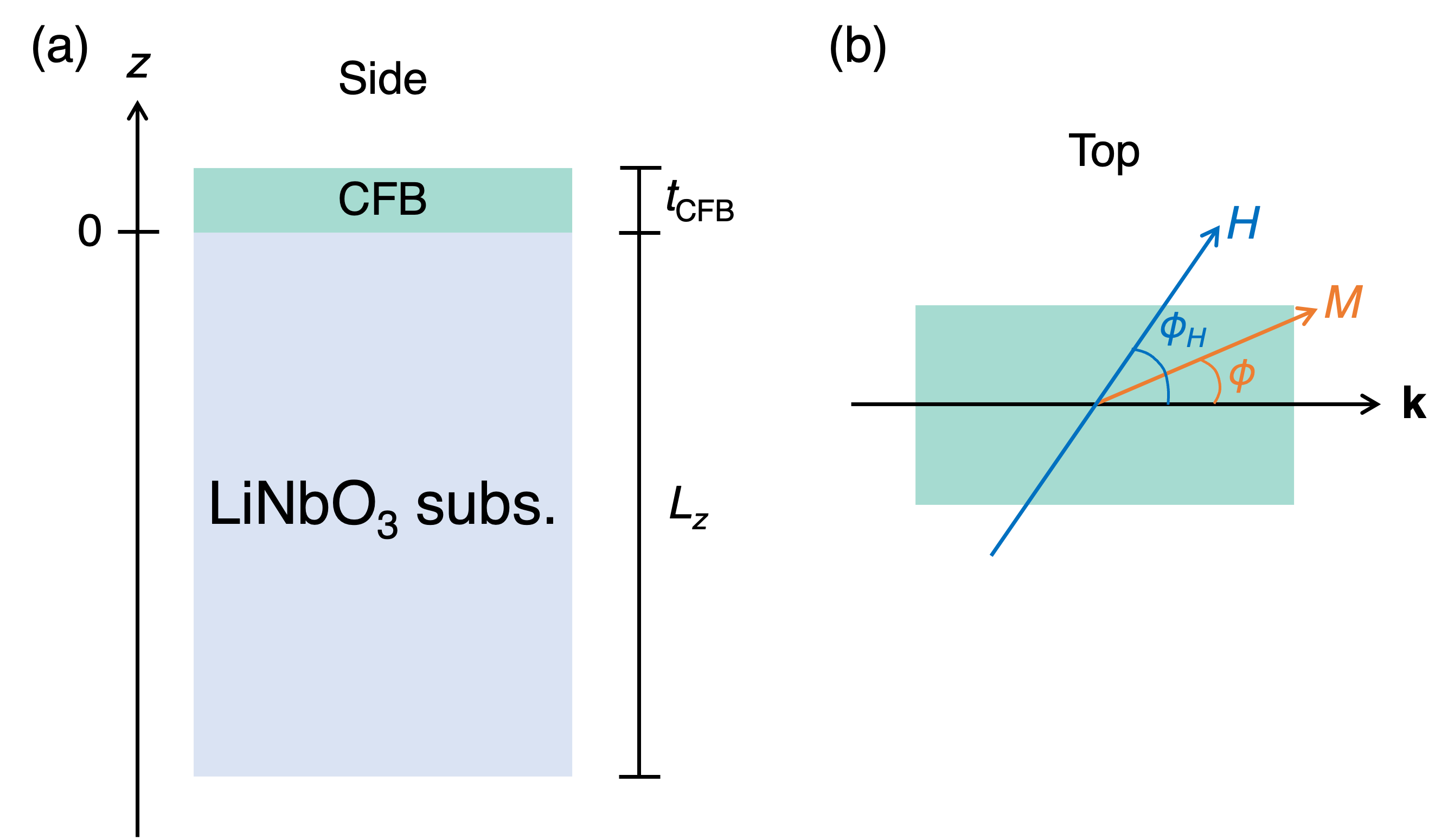}
\caption{\label{fig:model_setup} (a),(b) Setting for the magnon--phonon coupling model of
(a) the substrate and the CFB stack and (b) the in-plane coordinates.}
\end{figure}

Here, we consider a CFB layer with the thickness of $t_{\mathrm{CFB}}$
deposited on top of a LiNbO$_3$ substrate with the thickness of $L_z$,
as shown in Fig.~\ref{fig:model_setup}(a).
Note that, we omitted the Ti layers for simplicity.

Consider an in-plane magnetic field $H$ is applied externally
with the angle $\phi_H$ to the SAW wavevector $\mathbf{k}$, as shown in Fig.~\ref{fig:model_setup}(b).
The linearized equations of motions of the magnetization in units of magnetic field
in a cylindrical coordinate are~\cite{Yamamoto2020}
\begin{equation}\label{eq:LLG}
\begin{pmatrix}
-\mu_0 H_z + i\alpha \omega / |\gamma|	&	i\omega / \gamma	&	\mu_0 h_{\mathrm{eff}}^z \\
-i\omega / \gamma	&	-\mu_0 H_{\phi} + i\alpha \omega / |\gamma|	&	\mu_0 h_{\mathrm{eff}}^{\phi} \\
\mu_0 \overline{h_{\mathrm{eff}}^z} t_{\mathrm{CFB}}/L	&
\mu_0 \overline{h_{\mathrm{eff}}^{\phi}} t_{\mathrm{CFB}}/L	&
\frac{\rho}{M_s}\left(\frac{\omega^2}{k^2} - \frac{\omega_p^2}{k^2} + i\mu_v\omega \right)
\end{pmatrix}
\begin{pmatrix}
n_z \\ n_{\phi} \\ \varepsilon_{\mathbf{k}}
\end{pmatrix}
= \frac{\sigma}{M_s}
\begin{pmatrix}
0 \\ 0 \\ 1
\end{pmatrix},
\end{equation}
where $\sigma$ is the external stress, $\mu_v$ is the coefficient of viscosity,
$n_z$ and $n_{\phi}$ are components of the normalized spin vector,
$\omega_p$ the phonon frequency,
and $\varepsilon_{\mathbf{k}}$ is the amplitude of SAW strain.
The overline denotes complex conjugation.
Considering that the generated spin wave by the SAW has the same wavevector $\mathbf{k}$,
the static magnetic field components are given by
\begin{equation}\label{eq:Hz}
H_z = H \cos{(\phi - \phi_H)}
+ M_s \frac{1 - e^{-kt_{\mathrm{CFB}}}}{kt_{\mathrm{CFB}}}
+ H_K \cos^2{(\phi - \phi_K)} + \frac{A_{\mathrm{ex}}^2}{\mu_0 M_s},
\end{equation}
\begin{equation}\label{eq:Hphi}
H_{\phi} = H \cos{(\phi - \phi_H)}
+ M_s \left(1 - \frac{1 - e^{-kt_{\mathrm{CFB}}}}{kt_{\mathrm{CFB}}} \right) \sin^2{\phi}
+ H_K \cos{2(\phi - \phi_K)} + \frac{A_{\mathrm{ex}}^2}{\mu_0 M_s},
\end{equation}
where $\phi$ is the in-plane angle of the ground state magnetization $M$,
shown in Fig.~\ref{fig:model_setup}(b),
$M_s$ the saturation magnetization,
$\phi_K$ the in-plane angle of the easy axis,
$H_K$ the uniaxial anisotropy field,
$A_{\mathrm{ex}}$ the exchange stiffness,
and $k = |\mathbf{k}|$.
The terms on the right-hand side in Eqs. (\ref{eq:Hz}) and (\ref{eq:Hphi}) are, in the order from the left, the Zeeman, dipolar, anisotropy, and exchange fields.
The dynamical effective magnetic field components generated by acoustic waves
in the thin film limit read~\cite{Xu2020,Yamamoto2020,Yamamoto2022}
\begin{equation}\label{eq:h_z}
\mu_0 h^z_{\mathrm{eff}} = \frac{2c}{M_s} \nu_R \sqrt{kL_z} \xi_S^2 \sqrt{1 - \xi_P^2} \cos{\phi},
\end{equation}
\begin{equation}\label{eq:h_phi}
\mu_0 h^{\phi}_{\mathrm{eff}} = -i\frac{2}{M_s} \nu_R \sqrt{kL_z} \xi_S^2 \left(1 - \xi_S^2 \right)
\left(b_1 \sin{\phi}\cos{\phi} - \eta b_2 \cos{2\phi} \right),
\end{equation}
where $b_1$, $b_2$ are the cubic magnetoelastic coupling coefficients,
$c$ is the magnetorotation coupling coefficient,
normally $c = -\mu_0 M_s^2 / 2$ if it arises solely from the shape anisotropy,
and $\eta = \varepsilon_{xy} / \varepsilon_{xx}$ as described in Sec.~\ref{sec:strain}.
We approximated the SAW in our experiments to the Rayleigh wave,
which is commonly used for SAWs generated on a 128$^{\circ}$ Y-cut LiNbO$_3$ substrate~\cite{Weiler2011,Dreher2012}.
The parameters from the SAW normalization~\cite{Xu2020,Yamamoto2020,Yamamoto2022} are
$\nu_R \sim 1.6$, $\xi_P^2 = c_R^2 / c_P^2$, and $\xi_S^2 = c_R^2 / (2c_S^2)$,
where $c_P$, $c_S$, and $c_R$ are the longitudinal, transverse, and Rayleigh sound velocities, respectively.
The magnitude of these fields is a measure of the field amplitude when a strain of order unity is excited $|\varepsilon_{\mathbf{k}}| \sim 1$.
However, $|\varepsilon_{\mathbf{k}}|$ is the average strain across the whole elastic body,
which means for surface waves the strain near the surface would be much larger, which explains the enhancement factor $\sqrt{kL_z}$~\cite{Yamamoto2020,Yamamoto2022}.

The resonant frequencies of the hybridized modes can be calculated by turning off the dampings and setting the determinant of the matrix on the left-hand side of Eq.~(\ref{eq:LLG}) zero:
\begin{eqnarray}\label{eq:det}
0 &=& \left(\omega^2 - \gamma^2 \mu_0^2 H_z H_{\phi} \right)
	\left(\omega^2 - \omega_p^2 \right) \nonumber \\
	&& -\frac{\gamma \mu_0^2 M_s k^2}{\rho}\frac{t_{\mathrm{CFB}}}{L}
	\left\{\gamma \mu_0 \left(H_{\phi} \left|h_{\mathrm{eff}}^z\right|^2
	+ H_z \left|h_{\mathrm{eff}}^{\phi}\right|^2 \right)
	+ 2\omega \mathrm{Im}\left[h_{\mathrm{eff}}^z \overline{h_{\mathrm{eff}}^{\phi}} \right] \right\}.
\end{eqnarray}
Note that the last term in the curly bracket is related to the non-reciprocity.
Since this research mainly focuses on the coupling strength at $\phi = 0$,
where the non-reciprocity becomes zero~\cite{Xu2020}, we omit this term.
Then, Eq.~(\ref{eq:det}) yields
\begin{equation}
\omega^2 = \frac{\omega_m^2 + \omega_p^2}{2} \pm \frac{1}{2}
\sqrt{\left(\omega_m^2 - \omega_p^2 \right)^2 + \left(2\delta\omega_{\mathrm{bare}}^2\right)^2},
\end{equation}
where
\begin{equation}\label{eq:omega_bare}
\delta\omega_{\mathrm{bare}}^2 = 2\nu_R \xi_S^2
\sqrt{\frac{\gamma^2 \mu_0 k^3 t_{\mathrm{CFB}}}{\rho M_s}} \left\{\sqrt{H_z} (1 - \xi_S^2) \left(b_1 \sin{\phi}\cos{\phi} - \eta b_2\cos{2\phi} \right)
+ c \sqrt{H_{\phi}} \sqrt{1 - \xi_P^2} \cos{\phi}\right\}
\end{equation}
and $\omega_m = |\gamma|\mu_0\sqrt{H_z H_{\phi}}$.
To derive the value of the coupling $g$, we compute the level repulsion gap for $\omega_m = \omega_p$ at which
\begin{equation}
\omega = \sqrt{\omega_m^2 \pm \delta\omega_{\mathrm{bare}}^2}
\sim \omega_m \pm \frac{\delta\omega_{\mathrm{bare}}^2}{2\omega_m}
= \omega_m \pm g.
\end{equation}
Therefore,
\begin{equation}\label{eq:g}
g = \nu_R \xi_S^2
\sqrt{\frac{k^3 t_{\mathrm{CFB}}}{\rho M_s}}
\left\{\frac{(1 - \xi_S^2)}{\sqrt{\mu_0 H_{\phi}}} \left(b_1 \sin{\phi}\cos{\phi} - \eta b_2\cos{2\phi} \right)
+ c \sqrt{\frac{1 - \xi_P^2}{\mu_0 H_{z}}} \cos{\phi}\right\}.
\end{equation}

Since $\varepsilon_{xx}$, coupling with $b_1$, decreases abruptly away from the surface due to a change of sign of $u_x$~\cite{Babu2021},
we assume $\varepsilon_{xx}$ has the effective penetration depth $t_{\mathrm{eff}}$.
Then, Eq.~(\ref{eq:g}) is modified as
\begin{equation}\label{eq:g_teff}
g = \nu_R \xi_S^2
\sqrt{\frac{k^3}{\rho M_s}}
\left\{\frac{(1 - \xi_S^2)}{\sqrt{\mu_0 H_{\phi}}} \left(b_1 \sqrt{t_{\mathrm{eff}}} \sin{\phi}\cos{\phi} - \eta b_2 \sqrt{t_{\mathrm{CFB}}} \cos{2\phi} \right) + c \sqrt{t_{\mathrm{CFB}}} \sqrt{\frac{1 - \xi_P^2}{\mu_0 H_{z}}} \cos{\phi}\right\}.
\end{equation}
Especially, when $\phi_H = \phi = 0$ as Fig. 3 in the main text, it reduces to
\begin{equation}\label{eq:g_phi0}
g = \nu_R \xi_S^2
\sqrt{\frac{k^3 t_{\mathrm{CFB}}}{\rho M_s}}
\left\{\eta b \frac{1 - \xi_S^2}{\sqrt{\mu_0 H_{\phi}}} + c \sqrt{\frac{1 - \xi_P^2}{\mu_0 H_{z}}}\right\}
\end{equation}
with $b = b_1 = b_2$.
For the anticrossing fittings and calculations in Fig. 3 in the main text and Fig.~\ref{fig:anticrossing_sim}, we set the parameters as
$\nu_R = 1.6$, $\xi_S^2 = 0.41$, $\xi_P^2 = 0.32$, $\eta = 0.5$, and $\rho = 8,000$ kg/m$^3$.

\section{SAW transmission model under magnon--phonon coupling}
\begin{figure}[b]
\centering
\includegraphics[width=0.4\textwidth]{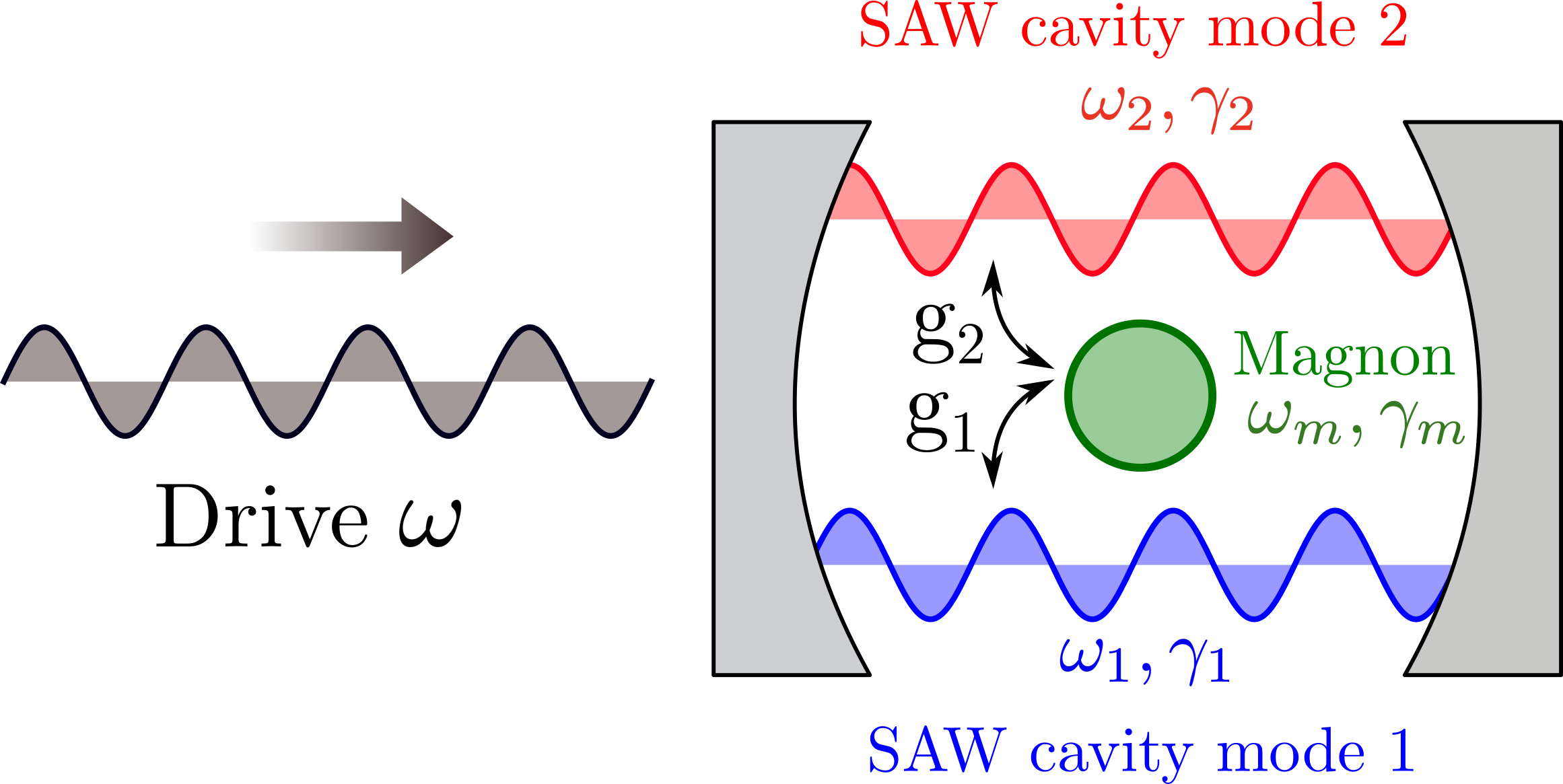}
\caption{\label{fig:transModel} \textbf{Sketch of the system for the transmission model.}}
\end{figure}

We model the system as shown in Fig.~\ref{fig:transModel}.
There are two acoustic SAW modes between the cavity mirrors---namely, the two modes observed in the cavity transmission in the absence of the magnet---of frequencies $\omega_1$ and $\omega_2$ and with intensity loss rates $\gamma_1$ and $\gamma_2$.
We assume that the intrinsic phonon loss is negligible,
so that the phonon loss is due to leakage through the mirrors outside the cavity
(for an extension including intrinsic loss, see below).
We also assume both mirrors have the same transmission coefficient.
In the absence of losses, each acoustic mode generates a displacement field
\begin{equation}
	\mathbf{u}_j(\mathbf{r},t) = \sqrt{A/(2\rho\omega_j)}\left[\mathbf{f}_j(\mathbf{r})e^{-i\omega_j t}a_j(t) + \text{c.c.}\right]
	\end{equation}
	with $\rho$ the material density, $\mathbf{f}_j(\mathbf{r})$ the normalized acoustic mode function, $a_j(t)$ the normalized amplitude and $A$ an arbitrary constant.
	Each acoustic mode is coupled to a single magnon mode (frequency $\omega_{mj}$, loss rate $\gamma_{mj}$) via the magnetoelastic interaction. In the absence of losses, the magnetization of each magnon mode can be decomposed in a similar way, 
	\begin{equation}
	\mathbf{m}_j(\mathbf{r},t) = \sqrt{A \vert\gamma\vert M_s}\left[\mathbf{h}_j(\mathbf{r})e^{-i\omega_{mj} t}s_j(t) + \text{c.c.}\right]
	\end{equation}
with $\gamma$ the gyromagnetic ratio, $M_s$ the saturation magnetization, $\mathbf{h}_j(\mathbf{r})$ the properly normalized mode function~\cite{CGBPRB} and $s_j(t)$ the mode amplitude. The losses of both magnon and acoustic modes originate from their coupling to fluctuating reservoirs. In the case of the magnon, we consider a single continuum reservoir labelled by an index $\omega$ and with normalized amplitudes $b_j(\omega,t)$. In the case of the acoustic modes, we consider two continuum baths describing the SAW continuum outside each of the cavity mirrors ($r$ = right, $l$ = left), and with respective amplitudes $a_\eta(\omega,t)$, where $\eta = r,l$.
The whole coupled dynamics obeys the following equations of motion,
	\begin{equation}\label{eom1}
		\frac{d a_j (t)}{dt} = -i\omega_j a_j(t) -i g^*_j s_j(t)
		-i\sum_\omega \sum_{\eta=r,l} \sqrt{\frac{\gamma_j}{8\pi}}a_\eta(\omega,t)
	\end{equation}
	\begin{equation}\label{eom2}
		\frac{d s_j (t)}{dt} = -i\omega_{mj} s_j(t) -i g_j a_j(t)
		-i\sum_\omega\sqrt{\frac{\gamma_{mj}}{4\pi}}b(\omega,t)
	\end{equation}
	\begin{equation}\label{eomAeta}
	\frac{d}{dt}a_\eta(\omega,t) = -i\omega a_\eta(\omega,t)-i\sum_j\sqrt{\frac{\gamma_j}{8\pi}}a_j(t)
	\end{equation}
	\begin{equation}
	\frac{d}{dt}b_j(\omega,t) = -i\omega b_j(\omega,t)-i\sum_j\sqrt{\frac{\gamma_{mj}}{4\pi}}s_j(t)
	\end{equation}
	with $g_j$ the magnetoelastic coupling rate between magnon mode $j$ and acoustic mode $j$.
	
	Let us now formally integrate the last two expressions in terms of a fixed point in the distant past $t_0\to -\infty$,
	\begin{equation}\label{aetat0}
		a_\eta(\omega,t) = e^{-i\omega(t-t_0)}a_\eta(\omega,t_0) -i\sum_j\sqrt{\frac{\gamma_j}{8\pi}}\int_{t_0}^t dsa_j(s) e^{-i\omega(t-s)}
	\end{equation}
	\begin{equation}
		b_j(\omega,t) = e^{-i\omega(t-t_0)}b_j(\omega,t_0) -i\sum_j\sqrt{\frac{\gamma_{mj}}{4\pi}}\int_{t_0}^t dss_j(s) e^{-i\omega(t-s)}
	\end{equation}
	and introduce them in Eqs.~\eqref{eom1}--\eqref{eom2}. After some algebra, we can cast the system evolution as
	\begin{equation}\label{EOMvvec}
		\frac{d}{dt}\left[
		\begin{array}{c}
			a_1  \\
			a_2 \\
			s_1 \\
			s_2
		\end{array}
		\right] \equiv \frac{d}{dt}\mathbf{v} = M\mathbf{v} - \mathbf{f}_{\rm in} \equiv  M\mathbf{v}- \left[
		\begin{array}{c}
			\sqrt{\gamma_1}\sum_\eta a_{\eta,\rm in}(t)  \\
			\sqrt{\gamma_2}\sum_\eta a_{\eta,\rm in}(t) \\
			\sqrt{\gamma_{m1}}b_{1\rm in}(t) \\
			\sqrt{\gamma_{m2}}b_{2\rm in}(t)
		\end{array}
		\right]
	\end{equation}
	with dynamical matrix
		\begin{equation}
		M= \left[
		\begin{array}{cccc}
		-i\omega_1-\gamma_1/2 & -\sqrt{\gamma_1\gamma_2}/2
		& -i g_1^* &0
		\\
		-\sqrt{\gamma_1\gamma_2}/2 & -i\omega_2-\gamma_2/2 & 0 & -i g_2^*
		\\
		-i g_1 & 0 & -i\omega_{m1}-\gamma_{m1}/2 & 0
		\\
		0 &-i g_2 &0 & -i\omega_{m2}-\gamma_{m2}/2 
		\end{array}
		\right]
		\end{equation}
and input field amplitudes are defined as
	\begin{equation}
	a_{\eta,\rm in}(t) \equiv \frac{i}{\sqrt{8\pi}}\sum_\omega e^{-i\omega (t-t_0)}a_\eta(\omega,t_0)
	\end{equation}
	\begin{equation}
	b_{j\rm in}(t) \equiv \frac{i}{\sqrt{4\pi}}\sum_\omega e^{-i\omega (t-t_0)}b_j(\omega,t_0).
	\end{equation}
	It is convenient to transform to the frequency domain using the convention
	\begin{equation}
	f(\omega) = \frac{1}{\sqrt{2\pi}}\int dt f(t) e^{i\omega t},
	\end{equation}
	where the system dynamics can be solved as
	\begin{equation}
	\mathbf{v}(\omega) = (M + i\omega)^{-1}\mathbf{f}_{\rm in}(\omega).
	\end{equation}

In order to compute transmission, we first derive input-output relations for the SAW bath field amplitudes. First, we use Eq.~\eqref{eomAeta} to solve for these amplitudes in terms of a time $t_1\to\infty$ in the distant future, obtaining
	\begin{equation}\label{aetat1}
		a_\eta(\omega,t) = e^{-i\omega(t-t_1)}a_\eta(\omega,t_1) +i\sum_j\sqrt{\frac{\gamma_j}{8\pi}}\int_{t}^{t_1} dsa_j(s) e^{-i\omega(t-s)}.
	\end{equation}
	We now introduce this expression into Eq.~\eqref{eom1} to obtain
	\begin{equation}\label{eom1inf}
		\frac{d a_j (t)}{dt} = -i\omega_j a_j(t) -i g_j^* s_j(t) +\frac{\gamma_j}{2}a_j(t)
		+\frac{\sqrt{\gamma_1\gamma_2}}{2}a_{\bar{j}}(t)
		-\sqrt{\gamma_j}\sum_\eta a_{\eta,\rm out}(t)
	\end{equation}
	where $\bar{j}$ is the opposite index of $j$ and with output amplitudes defined as
	\begin{equation}
	a_{\eta,\rm out}(t) \equiv \frac{i}{\sqrt{8\pi}}\sum_\omega e^{-i\omega (t-t_1)}a_\eta(\omega,t_1).
	\end{equation}
	Subtracting the corresponding component from Eq.~\eqref{EOMvvec}, namely
	\begin{equation}
		\frac{d a_j (t)}{dt} = -i\omega_j a_j(t) -i g_j^* s_j(t) -\frac{\gamma_j}{2}a_j(t)
		\\-\frac{\sqrt{\gamma_1\gamma_2}}{2}a_{\bar{j}}(t)
		-\sqrt{\gamma_j}\sum_\eta a_{\eta,\rm in}(t)
	\end{equation}
	we obtain the input-output relations
	\begin{equation}
	a_{\eta,\rm out}(t) = a_{\eta,\rm in}(t) + \sum_j\frac{\sqrt{\gamma_j}}{2}a_j(t).
	\end{equation}
	It is easy to see that the Fourier transform of the output amplitude is related to the transmission, as the occupation of the bath operators in the distant future can be cast as
	\begin{equation}
	\langle a_\eta^\dagger(\omega, t_1)a_\eta(\omega,t_1)\rangle = 4\left\langle\left[a_{\eta,\rm out}(\omega)\right]^\dagger
	a_{\eta,\rm out}(\omega)
	\right\rangle.
	\end{equation}
	Here, the brackets denote an ensemble average over any potential stochastic variables in the SAW bath fields, e.g. thermal fluctuations.

Assuming we drive the cavity from the left side, we can define the transmission as
	\begin{equation}
	T(\omega) \equiv \frac{\langle a_r^\dagger(\omega, t_1)a_r(\omega,t_1)\rangle}{\langle a_l^\dagger(\omega, t_0)a_l(\omega,t_0)\rangle}= \frac{\left\langle\left[a_{r,\rm out}(\omega)\right]^\dagger
		a_{r,\rm out}(\omega)
		\right\rangle}{\left\langle\left[a_{l,\rm in}(\omega)\right]^\dagger
		a_{l,\rm in}(\omega)
		\right\rangle}
	\end{equation}
	where we have used the easy-to-check identity
	\begin{equation}
	\left\langle\left[a_{\eta,\rm in}(\omega)\right]^\dagger
	a_{\eta',\rm in}(\omega)
	\right\rangle = \frac{1}{4}e^{i(\omega-\omega')t_0}\langle a_\eta^\dagger(\omega,t_0)a_{\eta'}(\omega',t_0)\rangle.
	\end{equation}
	We thus aim at computing $a_{r,\rm out}(\omega)$, given according to the input-output relation by
	\begin{equation}
	a_{r,\rm out}(\omega) = a_{r,\rm in}(\omega) + \sum_j\frac{\sqrt{\gamma_j}}{2}a_j(\omega).
	\end{equation}
	This expression can be written in a compact form as
	\begin{eqnarray}\label{aout}
		a_{r,\rm out}(\omega) = a_{r,\rm in}(\omega) &+& \mathbf{w}\cdot(M + i\omega)^{-1}\cdot\mathbf{w}\sum_\eta a_{\eta,\rm in}(\omega)\nonumber\\
		&+& \mathbf{w}\cdot(M + i\omega)^{-1}\cdot
		\left[
		\mathbf{e}_3\sqrt{\gamma_{m1} /2}b_{1\rm in}(\omega)
		+ \mathbf{e}_4\sqrt{\gamma_{m2} /2}b_{2\rm in}(\omega)
		\right]
	\end{eqnarray}
	where we have defined the vector
	\begin{equation}
	\mathbf{w} \equiv \left[\sqrt{\gamma_1/2},\sqrt{\gamma_2/2},0,0\right]^t.
	\end{equation}
	The initial values corresponding to our transmission experiment are
	\begin{equation}
	\langle a_\eta^\dagger(\omega,t_0)a_{\eta'}(\omega',t_0)\rangle = \delta_{\eta\eta'} \delta(\omega-\omega') \left[\bar{n}(\omega) + \delta_{\eta l} \vert\alpha\vert^2\right]
	\end{equation}
	\begin{equation}
	\langle b_j^\dagger(\omega,t_0)b_j(\omega',t_0)\rangle = \delta(\omega-\omega')\bar{n}(\omega)
	\end{equation}
	with $\bar{n}(\omega)$ the thermal distribution at the frequency $\omega$ and temperature $T$, that is, all the baths are on a thermal equilibrium state except for the left-propagating SAW bath, which on top of its thermal fluctuations contains the coherent drive amplitude $\alpha$ generated by the SAW generator. Using these values into Eq.~\eqref{aout} we can compute the transmission as
	\begin{eqnarray}
		T(\omega) &=& \left\vert \mathbf{w}\cdot(M + i\omega)^{-1}\cdot\mathbf{w} \right\vert^{2}\nonumber
		\\
		&+&\frac{\bar{n}(\omega)}{\bar{n}(\omega)+\vert\alpha\vert^2}\bigg[1 + \left\vert \mathbf{w}\cdot(M + i\omega)^{-1}\cdot\mathbf{w} \right\vert^{2} +
		\sum_{j=1,2}\gamma_{jm} \left\vert \mathbf{w}\cdot(M + i\omega)^{-1}\cdot\mathbf{e}_{j+2} \right\vert^{2}\bigg]
	\end{eqnarray}
	where the first term describes the transmission of the drive and the second term describes both the initial thermal fluctuations of the right SAW bath and the injection of thermal fluctuations from the system and the left reservoir. Assuming a drive amplitude much more intense than the thermal fluctuations, $\bar{n}(\omega) \ll \vert \alpha\vert^2$, we can approximate the transmission by the following analytical expression:
\begin{equation}\label{eq:T}
T(\omega) = \frac{1}{4} \left| \frac{A}{B}\right|^2,
\end{equation}
where
\begin{eqnarray}\label{eq:AandB}
A &=& \gamma_1 |g_2|^2 D_3(\omega) + \gamma_2 |g_1|^2 D_4(\omega)
+ D_3(\omega)D_4(\omega)(\gamma_2 D_1(\omega) + \gamma_1 D_2(\omega) + \gamma_1 \gamma_2),
\nonumber \\
B &=& D_1(\omega)D_2(\omega)D_3(\omega)D_4(\omega) + |g_1|^2 D_2(\omega) D_4(\omega)
+ |g_2|^2 D_1(\omega) D_3(\omega) \nonumber \\
&&- (\gamma_1 \gamma_2 / 4)D_3(\omega)D_4(\omega) + |g_1 g_2|^2,
\end{eqnarray}
	with
	\begin{equation}
	D_j(\omega) = i(\omega-\omega_1)-\gamma_1/2
	\end{equation}
	and with $g_j$ being in general complex numbers and with $j=1,2$ labeling the acoustic modes and $j=3,4$ labeling the magnon modes 1, 2, respectively.
The magnon frequency is $\omega_{mj} = \gamma\mu_0 \sqrt{H_z H_{\phi}}|_{k = k_j}$
with $k_j = \omega_j / v$ and the magnon linewidth is computed as
\begin{equation}
\gamma_{mj} = \frac{2\alpha\omega_{mj}}{\mu_0 |\gamma|} \frac{\partial \omega_{mj}}{\partial H}
= \alpha \mu_0 |\gamma| \left(H_z + H_{\phi} \right) \cos{(\phi - \phi_H)}.
\end{equation}

In the above expression, we have assumed that the only mechanism for phonon loss from the cavity is
the phonons leaking out from the reflectors, and as such
the transmission reaches a value of exactly 1 at resonance, i.e. the height of both peaks is equal.
This balance can be broken if we assume that the phonons have a total linewidth
$\gamma^{\mathrm{tot}}_j = \gamma_j = \Gamma_j$, where $\Gamma_j$ describes intrinsic losses.
In this case, the expression for the transmission is identical to Eqs. (\ref{eq:T})--(\ref{eq:AandB})
but with the substitution
\begin{equation}
D_{1,2}(\omega) = i(\omega - \omega_{1,2}) - \gamma^{\mathrm{tot}}_{1,2}/2.
\end{equation}
In addition, we have to consider that our acoustic cavity has a certain frequency range
around the main resonance $\omega_c$.
Finally, the transmission is expressed as
\begin{equation}\label{eq:T_filtered}
\tilde{T}(\omega) = T(\omega) \frac{1}{\sigma \sqrt{2\pi}} e^{-(\omega - \omega_c)^2 / (2\sigma^2)}.
\end{equation}

\subsection{In-plane magnetic field angular dependence of SAW-driven SWR}\label{sec:angDep}
\begin{figure}
\centering
\includegraphics[width=0.9\textwidth]{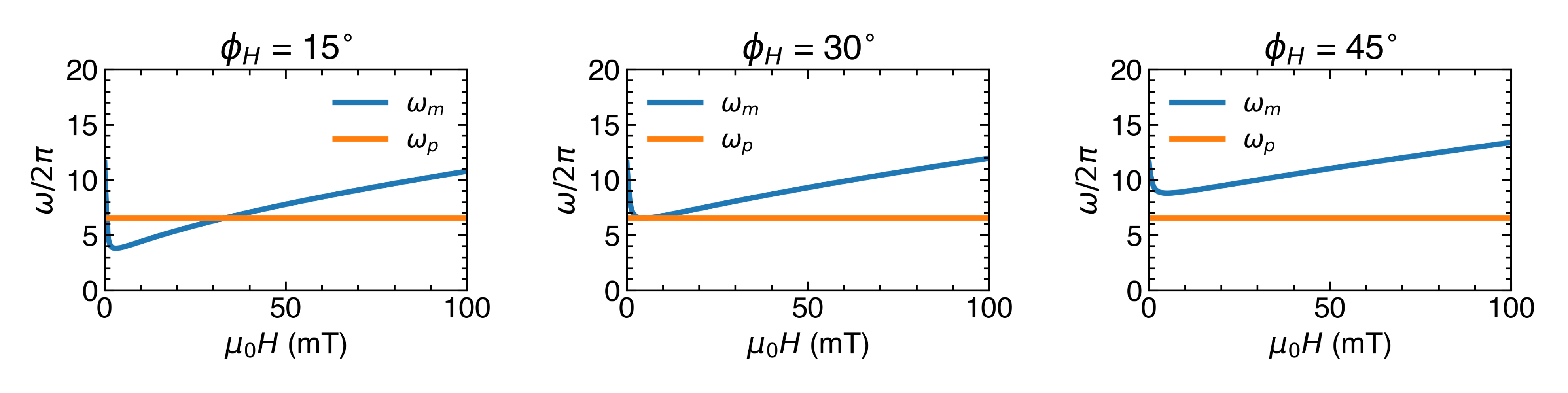}
\caption{\label{fig:SWR_angle} Magnon and phonon dispersions at the in-plane magnetic field angles $\phi_H = 15^{\circ}, 30^{\circ}$, and 45$^{\circ}$.
The orange lines show the fixed phonon frequency $\omega_p = 2\pi \times$ 6.58 GHz.
The blue curves show the SWR dispersion $\omega_m = |\gamma|\mu_0 \sqrt{H_z H_{\phi}}$}
\end{figure}

To calculate the in-plane magnetic field angle dependence of our magnon--phonon coupling,
we first calculated the ground state of the magnetization.
The free energy normalized by the saturation magnetization is
\begin{eqnarray}
G = &-&\mu_0 H \left\{\sin{\theta}\sin{\theta_H}\cos{(\phi - \phi_H)} + \cos{\theta}\cos{\theta_H} \right\}
\nonumber \\
&-& \mu_0 H_K \sin^2{\theta}\sin^2{\phi} + \frac{M_s}{2}\cos^2{\theta},
\end{eqnarray}
where $\theta$ ($\theta_H$) is the polar angle of the magnetization (magnetic field).
Note that the uniaxial magnetic anisotropy is aligned to the in-plane perpendicular to $\mathbf{k}$.
By taking the $\phi$ value to make it minimum as
$\partial G / \partial \phi |_{\theta=\theta_H=\pi/2} = 0$,
we get the angle of ground state magnetization.
Through substituting this $\phi$ value to Eq.~(\ref{eq:g_teff}) and the transmission model above,
we reproduced the angular dependence as shown in Fig. 2(d) in the main text with $\mu_0 H_K =$ 0.2 mT and $t_{\rm eff} =$ 2 nm.

Due to the magnetic anisotropy, the SWR dispersion of CFB with $k = 2\pi /$(600 nm) makes an additional crossing point with the phonon dispersion at $\phi_H < 30^{\circ}$, as shown in Fig.~\ref{fig:SWR_angle}.
This is the origin of the SAW absorption peaks close to $\mu_0H = 0$ at $\phi_H < 30^{\circ}$
in Figs. 2(b) and 2(c) in the main text and Fig.~\ref{fig:XandY}(a).

\subsection{Reproduced anticrossings by the SAW transmission model}
\begin{figure}
\centering
\includegraphics[width=0.55\textwidth]{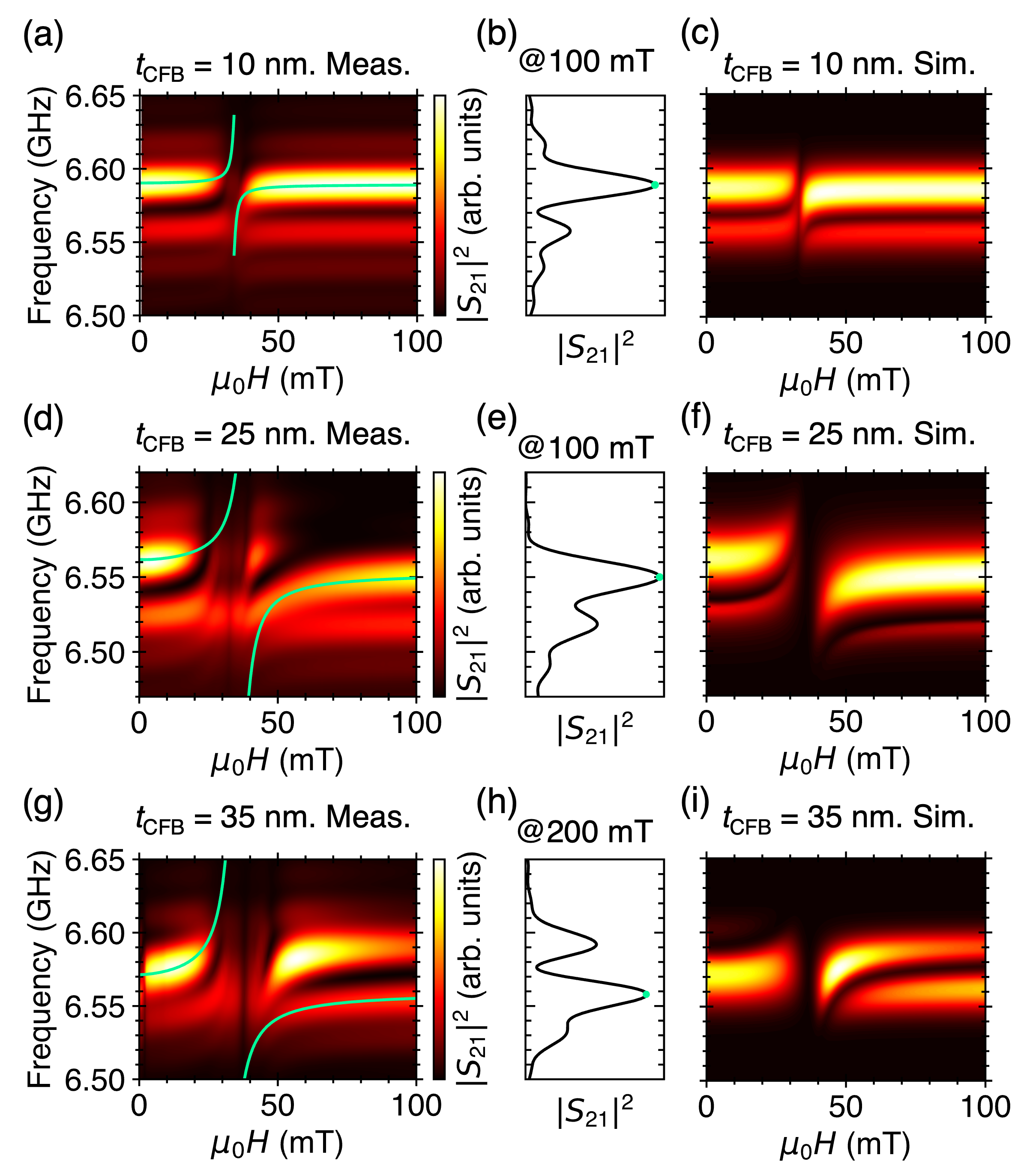}
\caption{\label{fig:anticrossing_sim} Measured and calculated SAW transmissions
when the external magnetic field is applied in the direction of SAW propagation; $\phi_H = 0$.
(a),(b) SAW transmission signal ($|S_{21}|^2$) of the sample
with $t_{\mathrm{CFB}} = 10$ nm under (a) various amplitudes of the magnetic field $\mu_0H$ 
and (b) $\mu_0H =$ 100 mT.
The green marker in (b) represents the local maximum used for the anticrossing fitting, shown as the green curves in (a).
(c) Calculated SAW transmission of the sample with $t_{\mathrm{CFB}} = 10$ nm
as a function of the frequency and $\mu_0H$.
(d),(e) $|S_{21}|^2$ of the sample
with $t_{\mathrm{CFB}} = 25$ nm under (d) various $\mu_0H$
and (e) $\mu_0H =$ 100 mT.
The green marker in (e) represents the local maximum used for the anticrossing fitting, shown as the green curves in (d).
(f) Calculated SAW transmission of the sample with $t_{\mathrm{CFB}} = 25$ nm
as a function of the frequency and $\mu_0H$.
(g),(h) $|S_{21}|^2$ of the sample
with $t_{\mathrm{CFB}} = 35$ nm under (g) various $\mu_0H$
and (h) $\mu_0H =$ 200 mT.
The green marker in (h) represents the local maximum used for the anticrossing fitting, shown as the green curves in (g).
(i) Calculated SAW transmission of the sample with $t_{\mathrm{CFB}} = 35$ nm
as a function of the frequency and $\mu_0H$.}
\end{figure}

Here, we show the measured SAW transmission spectra
under an external in-plane magnetic field at $\phi_H = 0$, the anticrossing fittings,
and reproduced SAW transmission spectra using Eq.~(\ref{eq:T_filtered})
not shown in Fig. 3 of the main text in Fig.~\ref{fig:anticrossing_sim}.

\section{Microwave ferromagnetic resonance measurements}
\begin{figure}
\centering
\includegraphics[width=0.5\textwidth]{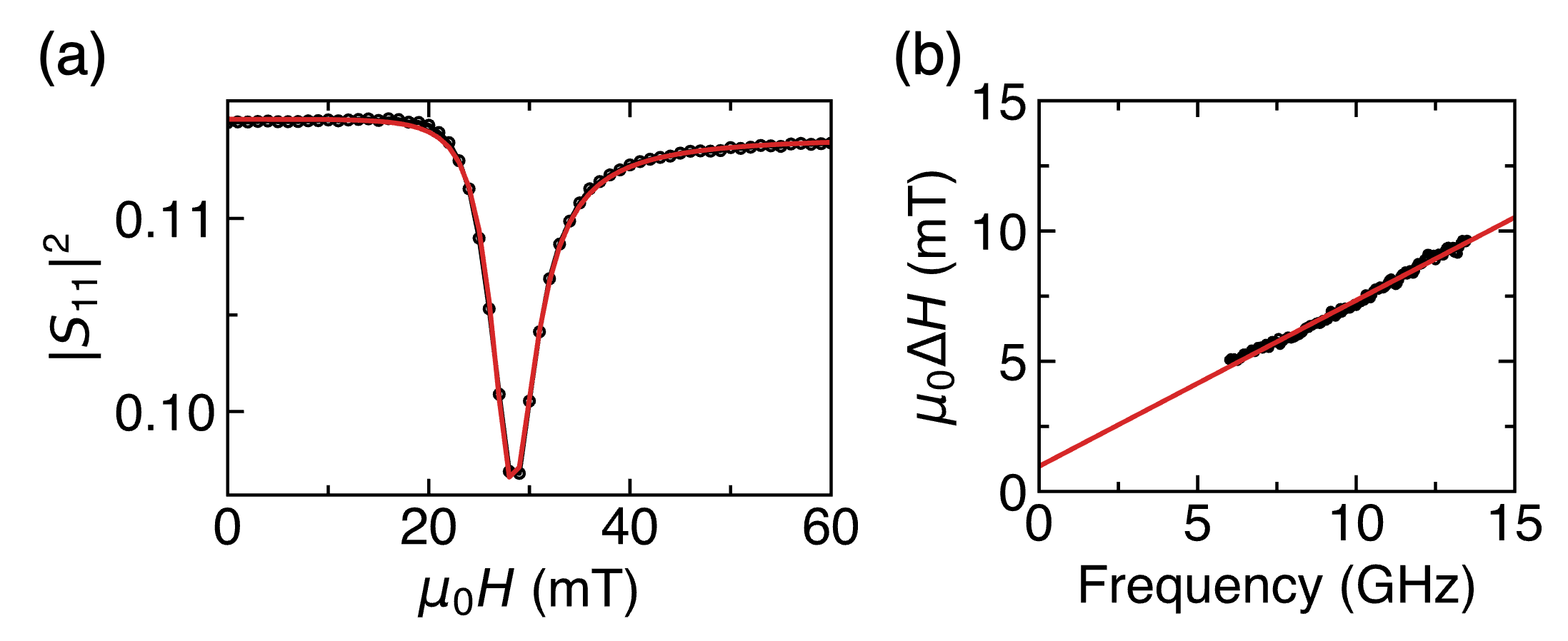}
\caption{\label{fig:FMR} Microwave FMR measurement of the CPW device with $t_{\mathrm{CFB}} = 20$ nm.
(a) SAW reflection at the frequency of 6.58 GHz as a function of the external magnetic field.
The markers and curve exhibit the measured data and the fitting, respectively.
(b), FMR linewidth as a function of frequency.
The markers and line exhibit the obtained values from the fittings
and the fitting with Eq.~(\ref{eq:linewidth}), respectively.}
\end{figure}

\begin{table}[htp]
\caption{Gilbert damping constant of each thickness of CFB.}\label{tab:Gilbert}
\begin{tabular}{ll}
\toprule
$t_{\mathrm{CFB}}$ (nm) &  $\alpha (\times 10^{-3})$ \\
\colrule
10 & 	9.21 $\pm$ 1.11 \\
20 &	8.97 $\pm$ 0.37 \\
25 &	11.8 $\pm$ 1.42 \\
30 &	9.44 $\pm$ 1.20 \\
35 &	8.20 $\pm$ 2.14 \\
\botrule
\end{tabular}
\end{table}

We measured the Gilbert damping constant $\alpha$ by ferromagnetic resonance (FMR)
to estimate the magnon relaxation rate.
A conventional coplanar waveguide (CPW) is used for microwave excitation.
The width and length of the Ti/CFB/Ti layers are 6 $\mu$m and 200 $\mu$m, respectively.
Signal lines of the CPW are deposited on Ti/CFB/Ti layers.
The width of the signal line of the CPW is 6 $\mu$m
and the distance between the signal line and ground line is 4 $\mu$m.
Note that the Ti/CFB/Ti layers are deposited at the same time as the samples
used for the magnon--phonon coupling experiments shown in the main text on LiNbO$_3$ substrates.
By applying an in-plane magnetic field perpendicular to the radio-frequency microwave field,
we observed the absorption of the microwave transmission $|S_{11}|^2$
as the example for the device with $t_{\mathrm{CFB}} = 20$ shown in Fig.~\ref{fig:FMR}(a).
Through Lorentzian fittings as shown as the curve in Fig.~\ref{fig:FMR}(a),
we obtain the FMR linewidth $\mu_0 \Delta H$
which has a well-known dependence upon the microwave frequency $f$ as
\begin{equation}\label{eq:linewidth}
\mu_0 \Delta H = \mu_0 \Delta H_0 + \frac{4\pi\alpha}{\gamma}f,
\end{equation}
where $\mu_0 \Delta H_0$ is the inhomogeneous linewidth.
Figure \ref{fig:FMR}(b) shows the obtained FMR linewidth as a function of the microwave frequency
and the fitting with Eq.~(\ref{eq:linewidth}) which yields the Gilbert damping constant.
The Gilbert damping constants estimated by the FMR linewidth fittings of all devices
are shown in Table \ref{tab:Gilbert}.

Additionally, note that the microwave absorption by FMR, as shown in Fig.~\ref{fig:FMR}(a),
we observe only a single absorption shape without any split features.
Therefore, we only observe anticrossing behaviors in CFB with our two-port SAW resonator devices.

\bibliography{02_sm}